%% file: main.tex
\documentclass[10pt,final,conference,letterpaper,twocolumn]{IEEEtran}
\usepackage{amsmath,amsthm,amsfonts,amssymb,bm}
\usepackage[table,xcdraw]{xcolor}
\usepackage{diagbox}
\usepackage{graphicx,marvosym}
\usepackage{xspace}
\usepackage{pifont}
\usepackage{multirow}
\usepackage{subcaption}
\usepackage{comment}
\usepackage{threeparttable}
\usepackage{amsmath}
\DeclareMathOperator*{\argmin}{argmin}
\newcommand{\argmax}{\operatornamewithlimits{argmax}}
\usepackage{booktabs}

\usepackage{algorithm}% http://ctan.org/pkg/algorithms
\usepackage[noend]{algpseudocode}% http://ctan.org/pkg/algorithmicx

\usepackage{enumitem}
\setlist{leftmargin=4mm}

\usepackage{tcolorbox}

\makeatletter
  \newcommand\figcaption{\def\@captype{figure}\caption}
  \newcommand\tabcaption{\def\@captype{table}\caption}
\makeatother

\newcommand{\tabincell}[2]{\begin{tabular}{@{}#1@{}}#2\end{tabular}}
\newcommand{\attackname}{\textsc{FakeBob}\xspace}

\usepackage{verbatim}
\usepackage{hyperref}
\usepackage{balance}

\usepackage{colortbl}

\usepackage{flexisym}

\usepackage{fancyhdr}
\newcommand{\distance}{2mm}
\begin{document}
\title{Who is Real Bob? Adversarial Attacks on \\Speaker Recognition Systems}

\author{\IEEEauthorblockN{
Guangke Chen\IEEEauthorrefmark{1}\IEEEauthorrefmark{2}\IEEEauthorrefmark{3},
Sen Chen\IEEEauthorrefmark{4}\IEEEauthorrefmark{6},
Lingling Fan\IEEEauthorrefmark{4},
Xiaoning Du\IEEEauthorrefmark{4},
Zhe Zhao\IEEEauthorrefmark{1},
Fu Song\IEEEauthorrefmark{1}\IEEEauthorrefmark{5}\textsuperscript{\Letter}
and Yang Liu\IEEEauthorrefmark{4}}
\IEEEauthorblockA{\IEEEauthorrefmark{1}ShanghaiTech University, \IEEEauthorrefmark{2}Shanghai Institute of Microsystem and Information Technology, Chinese Academy of Sciences}
\IEEEauthorblockA{\IEEEauthorrefmark{3}University of Chinese Academy of Sciences, \IEEEauthorrefmark{4}Nanyang Technological University}
\IEEEauthorblockA{\IEEEauthorrefmark{5}Shanghai Engineering Research Center of Intelligent Vision and Imaging, \IEEEauthorrefmark{6}Co-first Author}
 }

\maketitle
\pagestyle{plain}

\input{abstract}

\input{introduction}

\input{background}

\input{methodology}

\input{attack_approach}

\input{attack_experiment}

\input{discussion_future_work}

\input{related_work}

\input{conclusion}

\clearpage

\section*{Acknowledgments}
This research was partially supported by National Natural Science Foundation of China (NSFC) grants (No. 61532019 and No. 61761136011),
National Research Foundation (NRF) Singapore, Prime Ministers Office under its National Cybersecurity R\&D Program (Award No. NRF2014NCR-NCR001-30 and  No. NRF2018NCR-NCR005-0001), National Research Foundation (NRF) Singapore, National Satellite of Excellence in Trustworthy Software Systems under its Cybersecurity R\&D Program (Award No. NRF2018NCR-NSOE003-0001), and National Research Foundation Investigatorship Singapore  (Award No. NRF-NRFI06-2020-0001).

\balance 

\input{appendices}

\end{document}

%% file: abstract.tex
% !TeX root = ../main.tex

\newcommand{\nathaniel}[1]{{\leavevmode\color{blue}[#1]}}

\begin{abstract}
Speaker recognition (SR) is widely used in our daily life as a biometric authentication or identification mechanism. The popularity of SR brings in serious security concerns, as demonstrated by recent adversarial attacks. However, the impacts of such threats in the practical black-box setting are still open, since current attacks consider the white-box setting only.

In this paper, we conduct the first comprehensive and systematic study of the adversarial attacks on SR systems (SRSs) to understand their security weakness in the practical black-box setting. For this purpose, we propose an adversarial attack, named \attackname, to craft adversarial samples. Specifically, we formulate the adversarial sample generation as an optimization problem, incorporated with the confidence of adversarial samples and maximal distortion to balance between the strength and imperceptibility of adversarial voices. One key contribution is to propose a novel algorithm to estimate the score threshold, a feature in SRSs, and use it in the optimization problem to solve the optimization problem. We demonstrate that \attackname achieves {99\%} targeted attack success rate on both open-source and commercial systems. We further demonstrate that \attackname is also effective on both open-source and commercial systems when playing over the air in the physical world. Moreover, we have conducted a human study which reveals that it is hard for human to differentiate the speakers of the original and adversarial voices. Last but not least, we show that {four} promising defense methods for adversarial attack from the speech recognition domain become ineffective on SRSs against \attackname, which calls for more effective defense methods. We highlight that our study peeks into the security implications of adversarial attacks on SRSs, and realistically fosters to improve the security robustness of SRSs.
\end{abstract}

%% file: introduction.tex
% !TeX root = ../main.tex

\section{Introduction}\label{sec:intro}
Speaker recognition~\cite{kinnunen2010overview} is an {automated} technique
to identify a person from utterances which contain audio characteristics of the speaker.
Speaker recognition systems (SRSs) are ubiquitous in our daily life, ranging from biometric authentication~\cite{TDBank},
forensic tests~\cite{forensic-testing}, to personalized service on smart devices~\cite{7581588}.
Machine learning techniques are the mainstream method for implementing SRSs~\cite{RibasV19},
however, they are vulnerable to adversarial attacks {(e.g.,~\cite{BiggioCMNSLGR13,szegedy2013intriguing,lei2020pelican})}.
Hence, it is vital to understand the security implications of SRSs under adversarial attacks.

Though the success of adversarial attack on image recognition systems has been ported to the speech recognition systems in both
the white-box setting (e.g.,~\cite{carlini2018audio,yuan2018commandersong}) and black-box setting (e.g.,~\cite{taori2018targeted,KAM18}), relatively little research has been done on SRSs.
Essentially, the speech signal of an utterance consists of two major parts: the underlying text and the characteristics of the speaker. To improve {the performance}, speech recognition will minimize speaker-dependent variations to determine the underlying text or command, whereas speaker recognition will treat the phonetic variations as extraneous noise
to determine the source of the speech signal. Thus, adversarial attacks tailored to speech recognition systems may become ineffective on SRSs.

An adversarial attack on SRSs aims at crafting a sample from a voice uttered by some source speaker, so that it is misclassified as {one of the enrolled speakers} (untargeted attack) or {a target speaker} (targeted attack) by the system under attack, but still {correctly} recognized as the source speaker by ordinary users.
Though current adversarial attacks on SRSs~\cite{gong2017crafting,kreuk2018fooling} are promising,
they suffer from the following three limitations: (1) They are limited to the white-box setting by assuming the adversary has access to the information of the target SRS. Attacks in {a} more realistic black-box setting are still open.
(2) They only consider either the close-set identification task~\cite{gong2017crafting} that always classifies an arbitrary voice as one of the enrolled speakers~\cite{liu2014factor}, or the speaker verification task~\cite{kreuk2018fooling} that checks if an input voice is uttered by the unique enrolled speaker or not~\cite{reynolds2000speaker}.
Attacks on the open-set identification task~\cite{fortuna2005open}, which strictly subsumes both close-set identification and speaker verification, are still open. (3) They do not consider over-the-air attacks, hence it is unclear whether their attacks are still effective when playing over the air in the physical world.
Therefore, in this work, \emph{we investigate the adversarial attack on all the three tasks of SRSs in the practical black-box setting}, in an attempt to understand the security weakness of SRSs under adversarial attack in practice.

In this work, we focus on the  black-box setting, which assumes that the adversary can obtain at most the decision result and scores of the enrolled speakers for each input voice.
Hence attacks in the black-box setting is more practical yet more challenging than the existing white-box attacks~\cite{gong2017crafting,kreuk2018fooling}. We emphasize that the scoring and decision-making mechanisms of SRSs are different among recognition tasks~\cite{Homayoon11}. Particularly, we consider 40 attack scenarios (as demonstrated in Fig.~\ref{fig:attackscenarios}) in total differing in attack types (targeted vs. untargeted), attack channels (API vs. over the air),  genders of source and target speakers, and SR tasks (cf. \S\ref{subsec:threat}).
{We demonstrate our attack on 16 representative attack scenarios.}

To launch such a practical attack, two technical challenges need to be addressed: {(C1)} crafting adversarial samples as less imperceptible as possible in the black-box setting,
and {(C2)} making the attack practical, namely, adversarial samples are effective on an unknown SRS, even when playing over the air in the physical world. In this paper, we propose a practical black-box attack, named  \attackname,
which is able to {overcome} these challenges.

Specifically, we formulate the adversarial sample generation as an optimization problem.
The optimization objective is parameterized by a confidence parameter and the maximal distortion of noise amplitude in $L_\infty$ norm to balance between the strength and imperceptibility of adversarial voices,
instead of using noise model~\cite{yuan2018commandersong,ijcai2019-741,qin2019imperceptible}, due to its device- and background-dependency. {We also incorporate the score threshold, a key feature in SRSs, into the optimization problem.}
To solve the optimization problem,
we leverage an efficient gradient estimation algorithm, i.e., the natural evolution strategy (NES)~\cite{ilyas2018black}. However, even with the estimated gradients, none of the existing gradient-based white-box methods (e.g., \cite{goodfellow2014explaining,kurakin2016adversarial,yuan2018commandersong,carlini2017towards})
can be directly used to attack SRSs. {This is due to the score threshold mechanism,
where an attack fails if the predicated score is less than the threshold.}
To this end, we propose a novel algorithm to estimate the threshold,
based on which we leverage the {Basic Iterative Method (BIM)}~\cite{kurakin2016adversarial}  
with estimated gradients to solve  the  optimization  problem.

We evaluate \attackname for its attacking capabilities, on {3 SRSs (i.e., ivector-PLDA~\cite{dehak2010front}, GMM-UBM~\cite{reynolds2000speaker} and xvector-PLDA~\cite{SnyderGSMPK19})} in the popular open-source platform Kaldi~\cite{kaldigithub} in the research community and 2 commercial systems (i.e., Talentedsoft~\cite{Talentedsoft}
and Microsoft Azure~\cite{Azure}) which are proprietary without any
publicly available information about the internal design and implementations, hence completely black-box.
We evaluate \attackname using 16 representative attack scenarios (out of 40) based on the following five aspects:
(1) effectiveness/efficiency, (2) transferability, (3) practicability, (4) imperceptibility, and (5) robustness.

The results show that \attackname achieves 99\% targeted attack success rate (ASR) on all the
tasks of {ivector-PLDA, GMM-UBM and xvector-PLDA systems},
and 100\% ASR on the commercial system Talentedsoft within 2,500 queries on average (cf. \S\ref{sec:c-s-i-experiment}).
To demonstrate the transferability, %of \attackname,
we conduct a comprehensive evaluation of transferability attack
on {ivector-PLDA, GMM-UBM and xvector-PLDA systems} under cross-architecture, cross-dataset, and cross-parameter circumstances
and the commercial system Microsoft Azure.
%Consequently, we define different attacks to investigate the transferability of adversarial samples under cross-architecture, cross-dataset, and cross-parameter circumstances on both open-source and commercial product (i.e., Microsoft Azure~\cite{Azure}).
{\attackname is able to achieve 34\%-68\% transferability (attack success) rate except for the speaker verification of Microsoft Azure.
The transferability rate could be increased by crafting
high-confidence adversarial samples at the cost of increasing distortion.}
To further demonstrate the practicability and imperceptibility, %of \attackname,
we launch an over-the-air attack in the physical world and also conduct a human study on the Amazon Mechanical Turk platform~\cite{amazon_mturk}.
The results indicate that \attackname is effective when playing over the air in the
physical world against both the open-source systems and {the open-set identification task of} Microsoft Azure (cf. \S\ref{subsec:over})
and it is hard for humans to differentiate the speakers of the
original and adversarial voices (cf. \S\ref{sec:human-study-exper}).

Finally, we study four defense methods that are reported promising in speech recognition domain:
audio squeezing~\cite{yuan2018commandersong,yang2018characterizing},
local smoothing~\cite{yang2018characterizing}, quantization~\cite{yang2018characterizing} and temporal dependency-based detection~\cite{yang2018characterizing},
due to lacking of domain-specific defense solutions for adversarial attack on SRSs.
{The results demonstrate that these defense methods have limited effects on \attackname,  indicating that \attackname is a practical and powerful adversarial attack on SRSs.}

{Our study reveals that the security weakness of SRSs under black-box adversarial attacks.
This weakness could lead to lots of serious security implications.
For instance, the adversary could launch an adversarial attack (e.g., \attackname) to bypass biometric authentication on the financial transaction~\cite{TDBank,citi} and smart devices~\cite{7581588},
as well as high-security intelligent voice control systems~\cite{vehicles}} so that follow-up voice command attacks can be launched, e.g., CommanderSong~\cite{yuan2018commandersong} and hidden voice commands~\cite{carlini2016hidden}.
{For the voice-enabled cars using Dragon Drive~\cite{vehicles},
the attacker could bypass its voice biometrics
using \attackname so that command attacks can be launched to control
cars.}
Even for commercial systems, it is a significant threat under such a practical black-box adversarial attack,
which calls for more robust SRSs.
To shed further light, we discuss the potential mitigation and further attacks to understand the arm race in this topic.
In summary, our main contributions are:

\begin{itemize}
	\item {To our knowledge, this is the first study of targeted adversarial attacks on SRSs in the black-box setting.}
    Our attack is launched by not only using gradient estimation based methods,
    but also incorporating the score threshold into the adversarial sample generation.
    The proposed algorithm to estimate the score threshold is unique in SRSs.

   \item Our black-box attack addresses not only the speaker recognition tasks considered by existing white-box attacks but also the more general task, open-set identification, which has not been considered by previous adversarial attacks.	
	
\item {Our attack is demonstrated to be \emph{effective} on the popular open-source systems and commercial system Talentedsoft,
\emph{transferable} and \emph{practical} on the popular open-source systems and  the open-set
identification task of} Microsoft Azure even when playing over the air in the physical world.

\item Our attack is \emph{robust} against {four} potential defense methods which are reported very promising in speech recognition domain.
 Our study reveals the security implications of the adversarial attack on SRSs,
  which calls for more robust SRSs and more effective domain-specific defense methods.
\end{itemize}

For more information of \attackname, please refer to our website~\cite{fakebob} which includes voice samples and source code.

%% file: background.tex
% !TeX root = ../main.tex

\section{Background}\label{sec:background}
In this section, we introduce the preliminaries of  speaker recognition systems (SRSs)
and the threat model.

\subsection{Speaker Recognition System (SRS)}\label{sec:speaker_recognition}
Speaker recognition is an {automated} technique that allows
machines to recognize a person's identity based on his/her utterances using the  characteristics of the speaker.
It has been studied actively for four decades~\cite{Homayoon11},  % \cite{kinnunen2010overview,Homayoon11})
and {currently supported by a number of open-source platforms (e.g., Kaldi and MSR Identity~\cite{MSRIdentity})
% SIDEKIT~\cite{SIDEKIT}), % and ALIZ\'{E} 3.0~\cite{alizetoolkit}),
and commercial solutions (e.g., Microsoft Azure,  Amazon Alexa~\cite{Alexa}, Google home~\cite{googlehome}, Talentedsoft,  %iFLYTEK~\cite{iFLYTEK-VPR}, Tencent VPR~\cite{tencentVPR}
and SpeechPro VoiceKey~\cite{SpeechPro}).}
In addition, NIST actively organizes the Speaker Recognition Evaluation~\cite{nisteval} since 1996. %Speaker Recognition

\begin{figure}[t]
  \centering
  \includegraphics[width=0.48\textwidth]{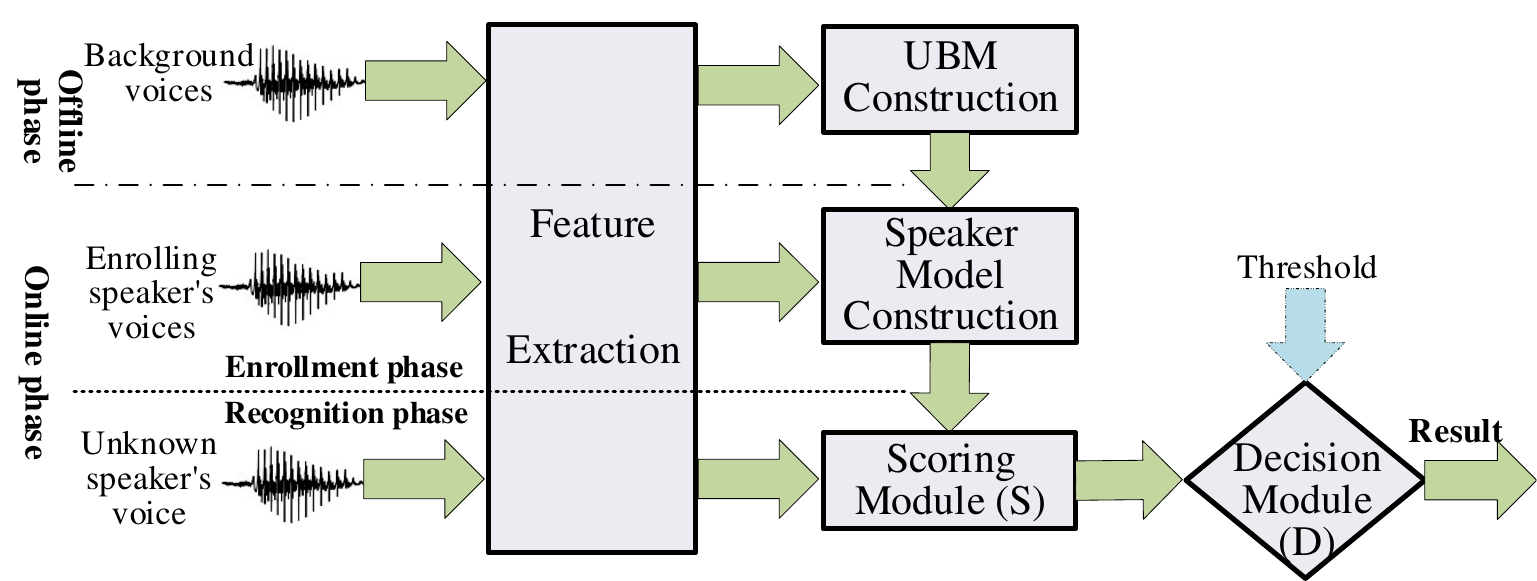}\\
  \vspace{-2mm}
  \caption{Overview of a typical SRS}\label{fig:overviewofSRS}
\end{figure}

\smallskip
\noindent \textbf{Overview of SRSs.}
Fig.~\ref{fig:overviewofSRS} shows an overview of a typical SRS,
which includes five key modules: Feature Extraction, Universal Background Model (UBM) Construction, Speaker Model Construction,
Scoring Module and Decision Module.
The top part is an offline phase,
while the lower two parts are an online phase composed of speaker enrollment and recognition phases.

In the offline phase, a UBM is trained using the acoustic feature vectors extracted from the background voices (i.e., voice training dataset) by the feature extraction module.
The UBM, intending to create a model of the average features of everyone in the dataset,
is widely used in the state-of-the-art SRSs
to enhance the robustness and improve efficiency~\cite{kinnunen2010overview}.
In the speaker enrollment phase, a speaker model is built using the
UBM and feature vectors of enrolling speaker's voices for each speaker.
During the speaker recognition phase, given an input voice $x$, the scores $S(x)$ of all the enrolled speakers are computed using the speaker models,
which will be emitted along with the decision $D(x)$ as the recognition result.

The feature extraction module converts a raw speech signal into acoustic feature vectors carrying  characteristics of the signal.
Various acoustic feature
extraction algorithms have been proposed
such as Mel-Frequency Cepstral Coefficients (MFCC)~\cite{MBE10},
Spectral Subband Centroid (SSC)~\cite{ThianSB04} and
Perceptual Linear Predictive (PLP)~\cite{Hermansky90}.
Among them, MFCC is the most popular one in
practice~\cite{kinnunen2010overview,Homayoon11}.

\smallskip
\noindent \textbf{Speaker recognition tasks.}
There are three common recognition tasks of SRSs: open-set identification (OSI)~\cite{fortuna2005open},
 close-set identification (CSI)~\cite{liu2014factor} and speaker verification (SV)~\cite{reynolds2000speaker}.

An OSI system allows multiple speakers
to be enrolled during the enrollment phase,
forming a speaker group $G$.
For an arbitrary input voice $x$, the system determines whether $x$ is uttered by one of the enrolled speakers
or none of them, according to the scores of all the enrolled speakers and a preset (score) threshold $\theta$.
Formally, suppose the speaker group $G$ has $n$ speakers $\{1,2,\cdots,n\}$,
the decision module outputs $D(x)$: % for an input voice $x$ as follows:
\begin{center}
$D(x)=\left\{
\begin{array}{ll}
\argmax\limits_{i\in G} \ [S(x)]_i, &\mbox{if } \max\limits_{i\in G} \ [S(x)]_i\geq \theta; \\
{\tt reject}, &\mbox{otherwise}.
\end{array}\right.$
\end{center}
where $[S(x)]_i$ for $i\in G$ denotes the score of the voice $x$ that is uttered by the speaker $i$.
Intuitively, the system classifies the input voice $x$ as the speaker $i$ if and only if the score $[S(x)]_i$ of the speaker $i$ is the largest one among all the enrolled speakers, and not less than the threshold $\theta$.
If the largest score is less than $\theta$, the system directly rejects the voice, namely,
it is not uttered by any of the enrolled speakers.

CSI and SV systems accomplish similar tasks as the OSI system, but with some special settings.
A CSI system never rejects any input voices, i.e., an input will always be classified as one of the enrolled speakers.
Whereas an SV system can have exactly \emph{one} enrolled speaker
and checks if an input voice is uttered by the enrolled speaker, i.e., either \emph{accept} or \emph{reject}.

\smallskip
\noindent \textbf{Text-Dependency.}
SRSs can be either text-dependent, where cooperative speakers are required to utter one of
pre-defined sentences,
or text-independent, where the speakers are allowed to speak anything.
The former achieves high accuracy on short utterances, but always requires a large amount utterances repeating the same sentence, thus it is \emph{only} used in the SV task.
The latter may require longer utterances to achieve high accuracy, but practically it is more versatile and can be used in all tasks (cf.~\cite{Homayoon11}).
Therefore, in this work, we mainly demonstrate our attack on text-independent SRSs.
% Remark that our attack is generic and could be applied to attack text-dependent SRSs.
% We leave this for further work.
%We also demonstrate attack on the text-dependent SV of Microsoft Azure.
%Thought it is difficult, our attack which has been demonstrated on Microsoft Azure.}
%, which has been demonstrated on Microsoft Azure.

\smallskip
\noindent{\bf SRS implementations.}
ivector-PLDA~\cite{dehak2010front,NFMCL19} is a mainstream method for implementing SRSs in both academia~\cite{kaldigithub,NidadavoluIVD19,LeeWK19} and industries~\cite{tencentVPR,Fosafer}. %changhong cumani2016fast
It achieves the state-of-the-art performance for all the speaker recognition tasks~\cite{7472652,Sremath2016areview}.
Another one is GMM-UBM based methods, which train a Gaussian mixture model (GMM)~\cite{reynolds2000speaker,ReynoldsR95} as UBM. Basically, GMM-UBM tends to provide comparative (or higher) accuracy on short
utterances~\cite{vestman2018speaker}.

Recently, deep neural network (DNN) becomes used in speech~\cite{AABCCCCCCD16} and speaker recognition (e.g., xvector-PLDA~\cite{SnyderGSMPK19}),
where speech recognition aims at determining the underlying text or command of the speech signal.
However, the major breakthroughs made by DNN-based methods reside in speech recognition;
for speaker recognition, ivector based methods still exhibit the state-of-the-art performance~\cite{RibasV19}.
Moreover, DNN-based methods usually rely on a much larger amount of training data, which could greatly increase the computational complexity compared
with ivector and GMM based methods~\cite{LiMJLZLCKZ17}, thus are not suitable for
off-line enrollment on client-side devices.
{We denote by ivector, GMM, and xvector the ivector-PLDA, GMM-UBM, and xvector-PLDA, respectively.}

\subsection{Threat Model}\label{subsec:threat}
We assume that the adversary intends to craft an adversarial sample from a voice uttered by some source
speaker, so that it is classified
as one of the enrolled speakers (untargeted attack) or the target speaker (targeted attack) by the SRS under attack, but is still recognized
as the source speaker by ordinary users.
% {Specifically, under targeted attack,
% one tries to mimic the voice of a targeted victim through an adversarial sample to unlock the smartphone, log in the application such as Facebook, and conduct illegal financial transactions successfully.
% Using untargeted attack, one can mimic the voice of anyone of enrolled speakers
% to bypass voice-based access control such as doors and IoT devices in a smart home which have enrolled multiple speakers. After bypassed authentication, follow-up voice command attacks (e.g., CommanderSong~\cite{yuan2018commandersong} and Hidden voice commands~\cite{carlini2016hidden})
% can be launched, e.g., on smart car~\cite{vehicles}.
% These attack scenarios are feasible in practice, e.g.,
% the victim cannot hear the adversarial voice
% or the source speaker is the attacker sitting in the presence of
% the victim when playing the adversarial voice.}

{
To deliberately attack the authentication of a target victim, we can compose adversarial voices, which mimic the voiceprint of the victim from the perspective of the SRSs.
Reasonably, the adversary can unlock the smartphones~\cite{Unlock-phone-vpr}, log into applications~\cite{wechat-vpr-logging},
% and conduct illegal financial transactions~\cite{TDBank}.
and conduct illegal financial transactions~\cite{TDBank}.
%Unlock the smartphone is also a typical scenario.
%Specifically, under targeted attack, one tries to mimic the voice of a targeted victim through an adversarial sample to unlock the smartphone, log in the application such as Facebook, and conduct illegal financial transactions successfully.
Under untargeted attack, we can manipulate voices to mimic the voiceprint of any one of enrolled speakers.
For example, we can bypass the voice-based access control such as iFLYTEK~\cite{iFLYTEK-VPR}, where multiple speakers are enrolled.
After bypassing the authentication, follow-up hidden voice command attacks (e.g., \cite{yuan2018commandersong,carlini2016hidden})
can be launched, e.g., on smart car with Dragon Drive~\cite{vehicles}.
These attack scenarios are practically feasible, for example, when the victim is not within the hearable distance of the adversarial voice,
or the attack voice does not raise the alertness of the victim due to the presence of other voice sources, either human or loudspeakers.}
%\sen{Not sure}

%{Specifically, for target attacks, typical attack scenarios include bypassing smartphone locks~\cite{TDBank,chen2017you}, login authentications, and financial transaction authentications, etc. In such scenarios, the attacker can mimic a voice of the targeted victim through an adversarial sample to further unlock the smartphone, log in the application such as Facebook, and conduct illegal financial transactions successfully.
%For untargeted attacks, voice control systems such as smart home is a typical attack scenario, where all family members enrolled in the IoT devices in a smart home. Therefore, the attacker can mimic the voice of anyone of them (i.e., without a specific target), then he can further control the IoT devices successfully. Such attacks can help launch the follow-up voice command attacks, e.g., CommanderSong~\cite{yuan2018commandersong} and Hidden voice commands~\cite{carlini2016hidden}.}
%\ling{@Prof. Song, please check this version.}

This paper focuses on the practical black-box setting where the adversary has
access \emph{only} to the recognition result (decision result and scores) of a target SRS for each test input, but not the
internal configurations or training/enrollment voices.
This black-box setting is feasible in practice, e.g.,
{the commercial systems Talentedsoft~\cite{Talentedsoft}, iFLYTEK, SinoVoice~\cite{SinoVoice}
and SpeakIn~\cite{SpeakIn}.
If the scores are not accessible (e.g., OSI task in the commercial system Microsoft Azure),
we can leverage transferability attacks. We assume the adversary has some voices of the target speakers
 to build a surrogate model, while these voices are not necessary the enrollment voices.
This is also feasible in practice as one can possibly record speeches of target speakers.
To our knowledge, the targeted black-box setting renders all previous adversarial attacks impractical on SRSs. Indeed, all the adversarial attacks on SRSs are white-box~\cite{gong2017crafting,kreuk2018fooling}
except for the concurrent work~\cite{ARGBWYST19}, which performs only untargeted attacks.}

\begin{figure}[t]
  \centering
  \includegraphics[width=0.5\textwidth]{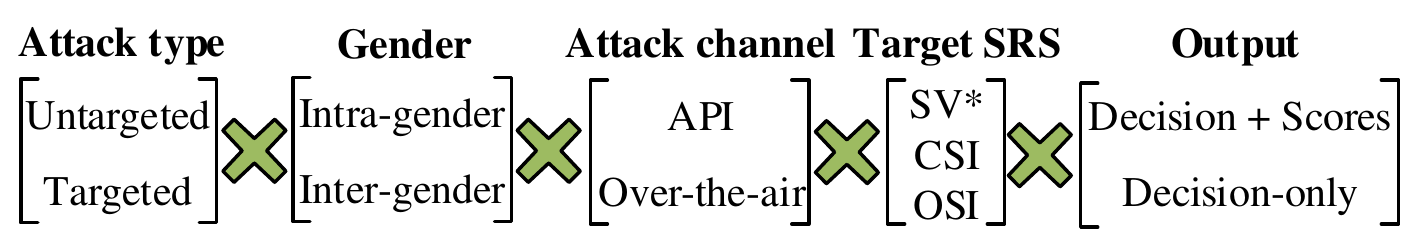}
  \vspace{-5mm}
  \caption{Attack scenarios, where
$*$ means that targeted and untargeted are the same on the SV task, as an SV system only has one enrolled speaker. }\label{fig:attackscenarios}
\end{figure}

Specifically, in our attack model, we consider five parameters: attack type (targeted vs. untargeted attack),
genders of speakers (inter-gender vs. intra-gender), attack channel (API vs. over-the-air),
speaker recognition task (OSI vs. CSI vs. SV) and output of the target SRS (decision and scores vs. decision-only) as shown in Fig.~\ref{fig:attackscenarios}.
%Untargeted attack aims to craft an adversarial sample that is recognized as one of the enrolled
%speakers by the system, while targeted attack aims to craft an adversarial sample that is recognized as the target speaker by the system.
Intra-gender (resp. inter-gender) means that the genders of the source and target speakers are the same (resp. different).
API attack assumes that the target SRS (e.g., Talentedsoft) provides an API interface to query,
while over-the-air means that attacks should be played over the air in the physical world.
%We simulate over-the-air attacks by playing voices via loudspeakers, recording
%the air-transmitted voices via receivers and then sending to the target system via API.
%To demonstrate the over-the-air attack, \fu{we use a built-in loudspeaker of a laptop (Lenovo) to play voices,
%and a built-in receiver of a mobile phone (OPPO) to record the air-transmitted voices (the distance between them is 0.5 meter~\cite{yuan2018commandersong}),
%finally, the recorded voices are fed to the system via API.}\fu{To be revised!}
Decision-only attack means that the target SRS (e.g., Microsoft Azure) only outputs decision result (i.e., the adversary can obtain the decision result $D(x)$), but not the scores of the enrolled speakers.
%(i.e., scores for each input voice $x$: $([S(x)]_i)_{i\in G}$ for OSI and CSI,
%and $S(x)$ for SV).
Therefore, targeted, inter-gender, over-the-air and decision-only attacks are the most practical yet the most challenging ones.
In summary, by counting all the possible combinations of the parameters in Fig.~\ref{fig:attackscenarios},
there are $48=2\times 2\times 2 \times 3\times 2$ attack scenarios.
Since targeted and untargeted attacks are the same on the SV task,
there are $40=48-2 \times 2\times 2$ attack scenarios.
{However, demonstrating all the 40 attack scenarios
requires huge engineering efforts, we design our experiments to cover 16 representative attack scenarios (cf. Appendix~\ref{sec:13attacks}).}

%% file: methodology.tex
% !TeX root = ../main.tex

\section{Methodology}\label{sec:overview}
In this section, we start with the motivations, then explain the design philosophy of our attack in black-box setting and the possible defenses,
finally present an overview of our attack.

\subsection{Motivation}
The research in this work is motivated by the following questions:
(Q1) How to launch an adversarial attack against all the tasks of SRSs in the practical black-box setting?
(Q2) Is it feasible to craft robust adversarial voices that are transferable to an unknown SRS under cross-architecture,
cross-dataset and cross-parameter circumstances, and commercial systems,  even when played over the air in the physical world?
(Q3) Is it possible to craft human-imperceptible adversarial voices that are difficult, or even impossible, to be noticed by ordinary users?
(Q4) If such an attack exists, can it be defended?

\subsection{Design Philosophy}\label{sec:designphi}
To address Q1, we investigate existing methods for black-box attacks on image/speech recognition systems,
i.e.,
surrogate model~\cite{papernot2017practical},
gradient estimation~\cite{chen2017zoo,ilyas2018black}
and genetic algorithm~\cite{SBBR16,alzantot2018genattack}.
Surrogate model methods are proved to be outperformed by gradient estimation methods~\cite{chen2017zoo}, hence are excluded.
For the other two methods:
it is known that natural evolution strategy (NES) based gradient estimation~\cite{ilyas2018black} requires much fewer queries than finite difference gradient estimation~\cite{chen2017zoo},
% it is known that natural evolution strategy (NES) based gradient estimation~\cite{ilyas2018black} requires much fewer queries than finite difference gradient estimation~\cite{ilyas2018black},
and particle swarm optimization (PSO) is proved to be more computationally efficient than other genetic algorithms~\cite{SBBR16,rios2013derivative}.
To this end, we conduct a comparison experiment on an OSI system using NES as a black-box gradient estimation technique
and PSO as a genetic algorithm.
The result shows that the NES-based gradient estimation method {obviously} outperforms the PSO-based one (cf. Appendix~\ref{sec:psocomparison}).
Therefore, we exploit the NES-based gradient estimation.

However, even with the estimated gradients,
none of the existing gradient based white-box methods (e.g., \cite{goodfellow2014explaining,kurakin2016adversarial,DLPSZHL18,MyMSTV17,yuan2018commandersong,qin2019imperceptible,ijcai2019-741,carlini2017towards})
can be directly used to attack SRSs.
This is due to the threshold $\theta$ which is used in the OSI and SV tasks, but not in
image/speech recognition.
As a result, these methods will fail to mislead SRSs
when the resulted score is less than $\theta$.
To solve this challenge, we incorporate the threshold $\theta$ into our adversarial sample generation
and propose a novel algorithm to estimate $\theta$  in the black-box setting.

Theoretically, the adversarial  samples crafted in the above way
are effective if directly fed as input to the target SRS via exposed API.
However, to launch a practical attack as in Q2, adversarial samples should be played
over the air in the physical world to interact with a SRS that may differ from the SRS on which adversarial samples are crafted.
To address Q2, we increase the strength of adversarial samples
and the range of noise amplitude, instead of using
noise model~\cite{yuan2018commandersong,ijcai2019-741,qin2019imperceptible}, due to its device- and background-dependency.
We have demonstrated that our approach is effective in transferability attack
even when playing over the air in the physical world.

To address Q3,
we should consider two aspects of the human-imperceptibility.
First, the adversarial samples should sound natural when listened by ordinary users.
Second, and more importantly, they should sound
as uttered by the same speaker of the original one.
As a first step towards addressing Q3, we add a constraint onto the perturbations
using $L_\infty$ norm, which restricts the maximal distortion
at each sample point of the audio signal.
We also conduct a real human study to illustrate the imperceptibility of our adversarial samples.

To address Q4, we should launch attacks on SRSs with defense methods. However, to our knowledge,
no defense solution exists for adversarial attacks on SRSs.
Therefore, we use four defense solutions for adversarial attacks on speech recognition systems:
audio squeezing~\cite{yuan2018commandersong,yang2018characterizing},
local smoothing~\cite{yang2018characterizing}, quantization~\cite{yang2018characterizing} and
temporal dependency detection~\cite{yang2018characterizing},
% temporal dependency-based detection~\cite{yang2018characterizing},
to defend against our attack.

\subsection{Overview of Our Attack: \attackname}
\begin{figure}[t]
	\centering
		\includegraphics[width=0.45\textwidth]{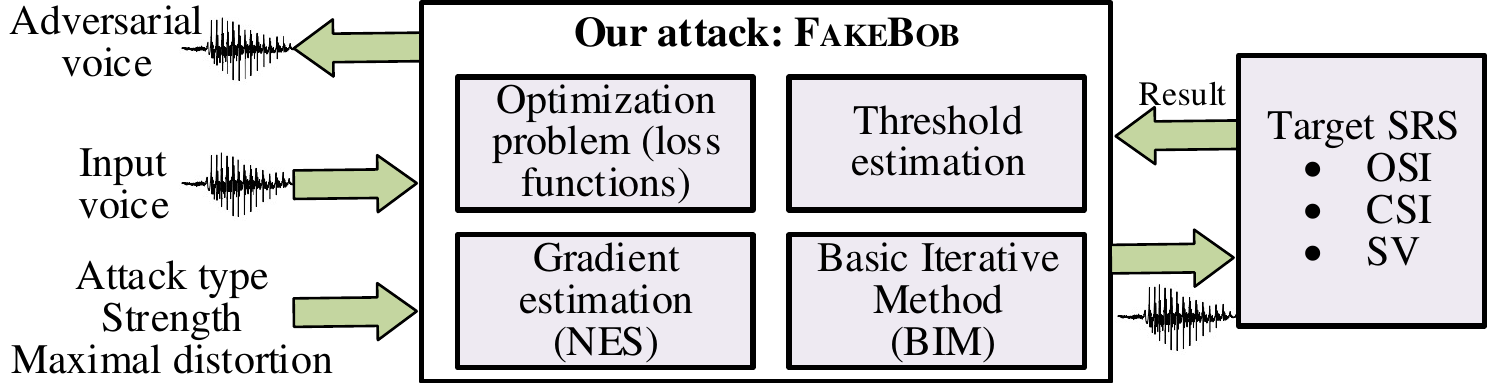}
        \vspace{-1mm}\caption{Overview of our attack: \attackname}
	    \vspace{-1mm}	\label{fig:overviewofattack}
\end{figure}

According to our design philosophy, in this section, we present an overview (shown in Fig.~\ref{fig:overviewofattack}) of our attack, named \attackname,
{addressing two technical
challenges (C1) and (C2) mentioned in \S\ref{sec:intro}.}
To address C1, we formulate adversarial sample generation as an optimization problem (cf. \S\ref{sec:problemformal}), for which specific loss functions are defined
for different attack types (i.e., targeted and untargeted) and tasks (i.e., OSI, CSI and SV) of SRSs (cf. \S\ref{attack_o_s_i}, \S\ref{attack_c_s_i} and \S\ref{attack_v}).
% To solve the optimization problem, we propose an approach
% by leveraging a novel algorithm to estimate the threshold, NES to estimate gradient and the BIM method with an estimated gradient.
To solve the optimization problem, we propose an approach
by leveraging a novel algorithm to estimate the threshold, NES to estimate gradient and the BIM method with the estimated gradients.
C2 is addressed by incorporating the maximal distortion ($L_\infty$ norm) of noise amplitude and
strength of adversarial samples into the optimization problem (cf. \S\ref{sec:problemformal},
\S\ref{attack_o_s_i}, \S\ref{attack_c_s_i} and \S\ref{attack_v}).
% C2 is addressed by incorporating the strength of adversarial samples
% and maximal distortion ($L_\infty$ norm) of noise amplitude into the loss functions (cf. Sections~\ref{attack_o_s_i}, \ref{attack_c_s_i} and \ref{attack_v}).

%% file: attack_approach.tex
% !TeX root = ../main.tex

\section{Our Attack: \attackname}\label{attack_approach}
In this section,
we elaborate on the techniques behind \attackname,
including the problem formulation
and attacks on OSI, CSI, and SV systems.

\subsection{Problem Formulation}\label{sec:problemformal}
Given an original voice, $x$, uttered by some source speaker,
the adversary aims at crafting an adversarial voice $\acute{x}=x+\delta$ by finding a perturbation $\delta$
such that (1) $\acute{x}$ is a valid voice~\cite{DZBS19}, (2) $\delta$ is as human-imperceptible as possible, and (3) the SRS under attack classifies
the voice $\acute{x}$ as one of the enrolled speakers or the target speaker.
To guarantee that the adversarial voice $\acute{x}$ is a valid voice,
which relies upon the audio file format (e.g., WAV, MP3 and AAC),
our attack \attackname first normalizes the amplitude value $x(i)$ of a voice $x$ at each sample point $i$ % in time
into the range $[-1,1]$, then crafts the perturbation $\delta$ to make sure
$-1\leq \acute{x}(i)=x(i)+\delta(i)\leq 1$, and finally
transforms $\acute{x}$ back to the audio file format which will be
fed to the target SRS.
Hereafter, we assume that the range of amplitude values is $[-1,1]$.
To be as human-imperceptible as possible,
our attack \attackname adapts $L_\infty$ norm to measure the similarity between the original and adversarial voices
and ensures that the $L_\infty$ distance $\|\acute{x}, x\|_\infty:=\max_i\{|\acute{x}(i)-x(i)|\}$
is less than the given maximal amplitude threshold $\epsilon$  of the perturbation,
where $i$ denotes sample point of the audio waveform.
To successfully fool the target SRS, we formalize the problem of
finding an adversarial voice $\acute{x}$ for a voice $x$ as the following constrained minimization problem:
\begin{equation}
\begin{array}{l}
  \argmin_\delta f(x+\delta)  \\
 \text{such that } \|x+\delta, x\|_\infty<\epsilon \mbox{ and } x+\delta\in {[-1,1]}^n
\end{array}
\label{eq:optimization-problem}
\end{equation}
where $f$ is a loss function. When $f$ is minimized, $x+\delta$ is recognized as the target speaker (targeted attack) or one of enrolled speakers (untargeted attack).
% where $f$ is a loss function when minimized, $x+\delta$ is recognized as the target speaker (targeted attack) or one of enrolled speakers (untargeted attack).
Our formulation is designed to be fast for minimizing the loss function rather than minimizing the perturbation $\delta$, as done in~\cite{goodfellow2014explaining,kurakin2016adversarial}.
Some studies,  e.g.,~\cite{carlini2017towards,szegedy2013intriguing}, formulate the problem to minimize both the loss function and perturbation.
It remains to define the loss function and algorithm to solve the optimization problem.
In the rest of this section, we mainly address them on the OSI system,
then adapt the solution to the CSI and SV systems.
% then adapt to the CSI and SV systems.

\begin{figure}[t]
	\centering
		\includegraphics[width=0.35\textwidth]{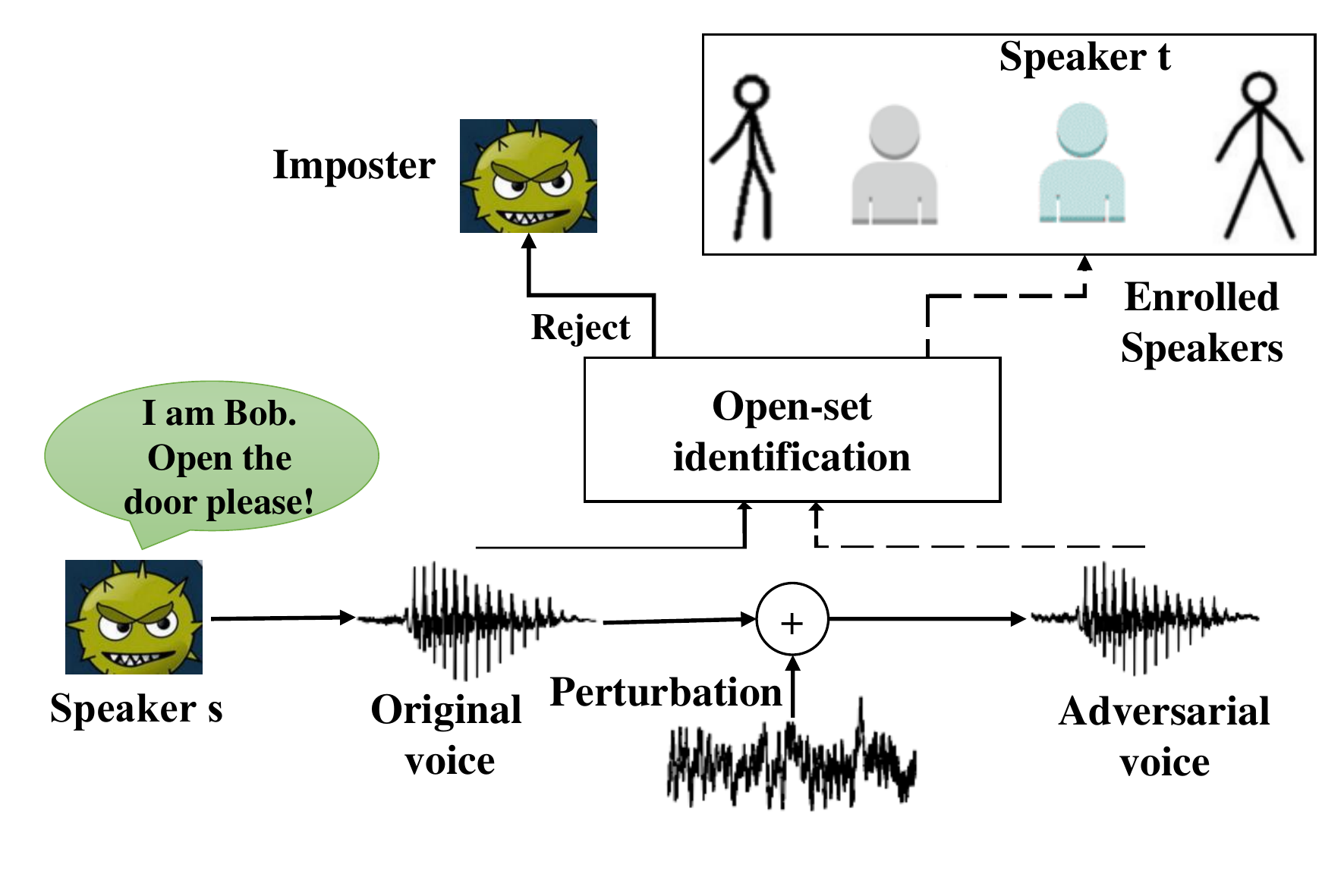}\vspace{-1em}
        \caption{Attack on OSI systems}
        \vspace{-1mm}
		\label{fig:attack o-s-i purpose}
\end{figure}

\subsection{Attack on OSI Systems}\label{attack_o_s_i}
As shown in Fig.~\ref{fig:attack o-s-i purpose},
to attack an OSI system, we want to
craft an adversarial voice $\acute{x}$ starting from
a voice $x$ uttered by some source speaker (i.e., $D(x)={\tt reject}$)
such that the voice $\acute{x}$ is classified
as the target speaker $t\in G=\{1,\cdots,n\}$  by the SRS,
i.e., $D(\acute{x})=t$.
We first present the loss function $f$ and then show how to solve the minimization problem.

\smallskip
\noindent{\bf Loss function $f$}. To launch a successful targeted attack on
an OSI system, the following two conditions
need to be satisfied \emph{simultaneously}:
the score $[S(x)]_t$ of the target speaker $t$  should be (1) the maximal one among all the enrolled speakers,
and (2) not less than the preset threshold $\theta$.
Therefore, the loss function $f$ for the target speaker $t$ is defined as follows:
\begin{equation}
\small
\label{eq:o_s_i_target_loss}
f(x)=\max\left\{(\max\{\theta,\max_{i\in G\setminus\{t\}}[S(x)]_i\}-[S(x)]_t),-\kappa\right\}
\end{equation}
{where the parameter $\kappa$, inspired by \cite{carlini2017towards}, intends to control
the strength of adversarial voices:
the larger the $\kappa$ is,
the more confidently the adversarial voice is recognized as the target speaker $t$ by the SRS.
% the larger $\kappa$,
% the stronger of the adversarial voice being recognized
% as the target speaker $t$.
This has been validated in \S\ref{sec:transferbility}.}

{Our loss function is similar to the one defined in~\cite{carlini2017towards},
but we also incorporate an additional threshold $\theta$.
% where we also incorporate the threshold $\theta$.
%The term $\max\{\theta,\max_{i\in G\setminus\{t\}}[S(x)]_i\}$
%selects the maximal value between the threshold $\theta$ and
%the scores of all the enrolled speakers except the target speaker $t$. (Remark that we will show how to estimate $\theta$ later.)
Considering $\kappa=0$, when $(\max\{\theta,\max_{i\in G\setminus\{t\}}[S(x)]_i\}-[S(x)]_t)$
is minimized, the score $[S(x)]_t$ of the target speaker $t$ will be maximized until it
exceeds the threshold $\theta$ and the scores of all other enrolled speakers.
Hence, the system recognizes the voice $x$ as the speaker $t$.
% Consider $\kappa=0$, when $(\max\{\theta,\max_{i\in G\setminus\{t\}}[S(x)]_i\}-[S(x)]_t)$
% is minimized, the score $[S(x)]_t$ of the target speaker $t$ will be maximized intended to
% exceed the threshold $\theta$ and the scores of all other enrolled speakers,
% thus the target system recognizes the voice $x$ as the target speaker $t$.
When $\kappa>0$,
instead of looking for a voice that just barely changes
the recognition result of $x$ to the  speaker $t$,
% instead of looking for a voice $x$ that just barely changes
% the recognition of $x$ as the target speaker $t$,
we want that the score $[S(x)]_t$ of the speaker $t$
is much larger than any other enrolled speakers and the threshold $\theta$.}

To launch an untargeted attack,
the loss function $f$ can be revised as follows:
\begin{equation}
\label{eq:o_s_i_untarget_loss}
f(x)=\max\{(\theta-\max_{i\in G}[S(x)]_i),-\kappa\}.
\end{equation}
Intuitively, we want to find a  perturbation $\delta$ such that the largest score of $x$
is at least $\kappa$ greater than the threshold $\theta$.

\smallskip\noindent{\bf Solving the optimization problem}.
To solve the optimization problem in Eq.~(\ref{eq:optimization-problem}),
we use NES as a gradient estimation technique and employ the BIM method with
the estimated gradients to craft adversarial examples.
Specifically,
% the BIM method begins by setting $\acute{x}_{i}=0$
the BIM method begins by setting $\acute{x}_{0}=x$
and then on the $i^{th}$ iteration,
\begin{center}
$\acute{x}_{i}={\tt clip}_{x,\epsilon}\{\acute{x}_{i-1}-\eta \cdot {\tt sign}(\nabla_x f(\acute{x}_{i-1}))\}$
\end{center}
where $\eta$ is a hyper-parameter indicating the learning rate, and the function ${\tt clip}_{x,\epsilon}(\acute{x})$, inspired by~\cite{kurakin2016adversarial}, performs per-sample clipping of the voice $\acute{x}$,
so the result will be in $L_\infty$ $\epsilon$-neighbourhood of the source voice $x$ and will be a valid voice
after being transformed back into the audio file format.
% after transforming back into the audio file format.
Formally, ${\tt clip}_{x,\epsilon}(\acute{x})=\max\{\min\{\acute{x},1,x+\epsilon\},-1,x-\epsilon\}$.

We compute the gradient $\nabla_x f(\acute{x}_{i-1})$ by leveraging NES,
which only depends on the recognition result.
In detail, on the $i^{th}$ iteration, we first create $m$ (must be even) Gaussian noises ($u_1,...,u_{m}$) and add them onto  $\acute{x}_{i-1}$, leading to
$m$ new voices
$\acute{x}_{i-1}^1,...,\acute{x}_{i-1}^m$,
where $\acute{x}_{i-1}^j=\acute{x}_{i-1}+\sigma\times u_{j}$ and $\sigma$ is the search variance of NES.
Note that $u_{j}=-u_{m+1-j}$ for $j=1,...,\frac{m}{2}$.
Then, we compute the loss values $f(\acute{x}_{i-1}^1),...,f(\acute{x}_{i-1}^m)$ by querying the target system ($m$ queries).
Next, the gradient $\nabla_x f(\acute{x}_{i-1})$ is approximated by computing
\begin{center}
$\frac{1}{m\times\sigma}\sum_{j=1}^{m}f(\acute{x}_{i-1}^j)\times u_{j}.$
\end{center}
In our experiments, $m=50$ and $\sigma=1e-3$.
Finally, we compute ${\tt sign}(\nabla_x f(\acute{x}_{i-1}))$,
a vector over the domain $\{-1,0,1\}$,
by applying element-wise ${\tt sign}$ mathematical operation to the gradient vector $\frac{1}{m\times\sigma}\sum_{j=1}^{m}f(\acute{x}_{i-1}^j)\times u_{j}$.

However, the BIM method with
the estimated gradients alone is not sufficient to construct adversarial samples in the black-box setting, due to the fact that
the adversary has no access to the threshold $\theta$ used in the loss function $f$.
To solve this problem, we present a novel algorithm for estimating $\theta$.

\smallskip
\noindent
{\bf Estimating the threshold $\theta$}.
To estimate the threshold $\theta$,
the main technical challenge is that the estimated threshold $\acute{\theta}$ should be no less than
$\theta$ in order to launch a successful attack,
but should not exceed $\theta$ too much, otherwise, the attack cost might become too expensive.
{Therefore, the goal is to compute a small $\acute{\theta}$ such that $\acute{\theta}\geq \theta$.
To achieve this goal,} we propose a novel approach as shown in Algorithm~\ref{al:threshold estimation2}.
Given an OSI system with the
scoring $S$ and decision $D$ modules,
and an arbitrary voice $x$ such that $D(x)={\tt reject}$, i.e.,
$x$ is uttered by an imposter,
%$x$ is uttered by some source speaker,
{Algorithm~\ref{al:threshold estimation2} outputs $\acute{\theta}$
such that $\acute{\theta}\geq \theta$}.

\begin{algorithm}[t]\footnotesize
	\caption{Threshold Estimation Algorithm}
	\label{al:threshold estimation2}
	%\resizebox{0.5\textwidth}{!}{%
	\begin{algorithmic}[1]
		\Require \begin{tabular}{ll}
             The target OSI system with scoring $S$ and decision $D$ modules \\
             An arbitrary voice $x$ such that $D(x)={\tt reject}$\\
           \end{tabular}
%		\Require An arbitrary voice $x$ such that $D(x)={\tt reject}$, the target OSI system with scoring $S$ and decision $D$ modules
		\Ensure Estimated threshold $\acute{\theta}$
		\State $\acute{\theta} \gets \max_{i\in G}[S(x)]_i$; \Comment{\textit{initial threshold}}
		\State $\Delta\gets |\frac{\acute{\theta}}{10}|$;  \Comment{\textit{the search step}}
		%\hfill// \textit{the search step}
		\State $\acute{x} \gets x$;
		%	\State $i=0$
		\While{True}
		\State{$\acute{\theta}\gets \acute{\theta}+\Delta$;}
        \State {$f'\gets\lambda x. \max\{\acute{\theta}-\max_{i\in G}[S(x)]_i,-\kappa\}$;} \Comment{\textit{loss function}}				\While{True}
		\State $\acute{x}\gets {\tt clip}_{x,\epsilon}\{\acute{x}-\eta \cdot {\tt sign}(\nabla_x f'(\acute{x}) )\}$;  \Comment{\textit{craft sample using $f'$}}
		%\hfill// craft a sample using $f'$
		\If{\textit{$D(\acute{x})\neq {\tt reject}$}};
		\Comment{\textit{$\max_{i \in G}[S(\acute{x})]_i\geq \theta$}}%\hfill// $\max_{i \in G}[S(\acute{x})]_i\geq \theta$
		\State \Return{$\max_{i \in G}[S(\acute{x})]_i$};
		\EndIf
		\If{$\max_{i\in G}[S(\acute{x})]_i\geq \acute{\theta}$} \textbf{break};
		\EndIf
		\EndWhile
		\EndWhile
	\end{algorithmic}%
	%}
\end{algorithm}

{In detail, Algorithm~\ref{al:threshold estimation2} first computes
the maximal score $\acute{\theta}=\max_{i\in G}[S(x)]_i$ of the voice $x$ by querying the system (line 1).
Since $D(x)={\tt reject}$, we can know $\acute{\theta}<\theta$.
At Line 2, we initialize the search step $\Delta=|\frac{\acute{\theta}}{10}|$, which will be used to
estimate the desired threshold $\acute{\theta}$.
$|\frac{\acute{\theta}}{10}|$ is chosen as a tradeoff between the precision of $\acute{\theta}$ and efficiency of the algorithm.}
{The outer-while loop (Lines 4-11) iteratively computes a new
candidate $\acute{\theta}$ by adding $\Delta$ onto it (Line 5)
and computes the function $f'=\lambda x. \max\{\acute{\theta}-\max_{i\in G}[S(x)]_i,-\kappa\}$ (Line 6).
 $f'$ indeed is the loss function for untargeted attack in Eq.~(\ref{eq:o_s_i_untarget_loss}), in which
$\theta$ is replaced by the candidate $\acute{\theta}$.
The function $f'$ will be used to craft samples in the inner-while loop (Lines 7-11).}
{For each candidate $\acute{\theta}$,
the inner-while loop (Lines 7-11) iteratively computes samples $\acute{x}$ by querying the target system
until the target system recognizes
$\acute{x}$ as some enrolled speaker (Line 9) or the maximal score of $\acute{x}$ is no less than $\acute{\theta}$ (Line 11).
If  $\acute{x}$ is recognized as some enrolled speaker (Line 9), then
Algorithm~\ref{al:threshold estimation2} terminates and returns the maximal score of $\acute{x}$ (Line 10),
as $\max_{i\in G}[S(\acute{x})]_i\geq \theta$ is the desired threshold.
%it returns the maximal score of $\acute{x}$ (Line 10), as $\max_{i\in G}[S(\acute{x})]_i\geq \theta$ is the desired threshold.
If the maximal score of $\acute{x}$ is no less than $\acute{\theta}$ (Line 11),
we restart the outer-while loop.}

One may notice that Algorithm~\ref{al:threshold estimation2} will not terminate
when \textit{$D(\acute{x})$}
is always equal to ${\tt reject}$.
In our experiments, this never happens (cf. \S\ref{sec:experiment}).
Furthermore, it
estimates a very close value to the actual threshold.
%Furthermore. it
%estimates a very close value of the actual threshold.
Remark that the actual threshold $\theta$, obtained from the open-source SRS, is used to evaluate
the performance of Algorithm~\ref{al:threshold estimation2} \emph{only}.

\subsection{Attack on CSI Systems}\label{attack_c_s_i}

%A CSI system always classifier an input voice as one of the enrolled speakers.
A CSI system always classifies an input voice as one of the enrolled speakers.
Therefore, we can adapt the attack on the OSI systems
by ignoring the threshold $\theta$.
Specifically, the loss function for targeted attack on CSI systems with the target speaker $t\in G$ is defined as:
%Specifically, the loss function for targeted attacking on CSI systems with the target speaker $t\in G$ is defined as:
\begin{center}
\small
$f(x)=\max\left\{(\max_{i\in G\setminus \{t\}}[S(x)]_i-[S(x)]_t),-\kappa\right\}$
\end{center}
Intuitively, we want to find some small perturbation $\delta$
such that the score of the speaker $t$ is the largest one among all the enrolled speakers,
and $[S(x)]_t$ is at least $\kappa$ greater than the second-largest score.
%and, $[S(x)]_t$ is at least $\kappa$ greater than the second-largest score.

Similarly, the loss function for untargeted attack on CSI systems is defined as:
%Similarly, the loss function for untargeted attacking on CSI systems is defined as:
\begin{center}
\small $f(x)=\max\{([S(x)]_m-\max_{i\in G\setminus\{m\}}[S(x)]_i),-\kappa\}$
\end{center}
where $m$ denotes the true speaker of the original voice.
%where $m$ denotes the speaker of the original voice.
Intuitively, we want to find some small perturbation $\delta$
such that the largest score among other enrolled speakers is at least $\kappa$ greater than the score of the speaker $m$.
%such that the largest score of other enrolled speakers is at least $\kappa$ greater than the score of the speaker $m$.

\subsection{Attack on SV Systems}\label{attack_v}
An SV system has exactly one enrolled speaker
and checks if an input voice is uttered by the enrolled speaker or not.
Thus, we can adapt the attack on OSI systems
by assuming the speaker group $G$ is a singleton set.
Specifically, the loss function for attacking SV systems is defined as:
\begin{center}\small
$f(x)=\max\{\theta-S(x),-\kappa\}$
\end{center}
Intuitively, we want to find a small perturbation $\delta$ such that
the score of $x$ being recognized as the enrolled speaker is at least $\kappa$ greater than the threshold $\theta$.
We remark that the threshold estimation algorithm for SV systems should be
revised by replacing the loss function $f'$ at {Line 6} in Algorithm~\ref{al:threshold estimation2} with
the following function:
%revised by replacing the loss function $f'$ at Line 7 in Algorithm~\ref{al:threshold estimation2} with
%the following function:
%\begin{center}
$f'=\lambda x. \max\{\acute{\theta}- S(x), -\kappa \}$.
%\end{center}

%% file: attack_experiment.tex
% !TeX root = ../main.tex

\begin{table}[t]\setlength{\tabcolsep}{3pt}
\centering
	\caption{Dataset for experiments}
	\vspace{-2mm}
	\label{tab:dataset-info}
  \begin{tabular}{c|c|l}
			\toprule
			\textbf{Datasets} & \textbf{\#Speaker} &\textbf{Details} \\ \midrule
			\tabincell{c}{{\bf Train-1}\\ {\bf Set} } &
			7,273
			& \tabincell{l}{Part of {\bf VoxCeleb1}~\cite{Nagrani17} and whole {\bf VoxCeleb2} \\ \cite{Chung18b} used for training ivector and GMM} \\ \midrule
			\tabincell{c}{{\bf Train-2}\\ {\bf Set}}   &
			2,411
			& \tabincell{l}{Part of {\bf LibriSpeech}~\cite{7178964} \\ used for training system C in transferability} \\ \midrule
			\tabincell{c}{{\bf Test}\\ {\bf Speaker}\\{\bf Set}} & 5 & \tabincell{l}{5 speakers from {\bf LibriSpeech} \\ 3 female and 2 male, 5 voices per speaker,\\ voices range from 3 to 4 seconds}\\ \midrule
			\tabincell{c}{{\bf Imposter}\\ {\bf Speaker}\\{\bf Set}}  & 4& \tabincell{l}{Another 4 speakers from {\bf LibriSpeech} \\ 2 female and 2 male, 5 voices per speaker, \\voices range from 2 to 14 seconds}  \\ \bottomrule
	\end{tabular}%
\end{table}

\section{Attack Evaluation}\label{sec:experiment}
We evaluate \attackname for its attacking capabilities  based on  the  following  five  aspects: {effectiveness/efficiency}, {transferability}, {practicability}, {imperceptibility}, and {robustness}.
	
\subsection{{Dataset and Experiment Design}}\label{sec:kaldi_dataset}
\noindent{\bf Dataset.} We mainly use three widely used datasets: VoxCeleb1, VoxCeleb2, and LibriSpeech (cf. Table~\ref{tab:dataset-info}).
To demonstrate our attack,
we target the ivector and GMM systems from the popular open-source platform Kaldi,
having 7,631 stars and 3,418 forks on Github~\cite{kaldigithub}.
The UBM model is trained using the \emph{Train-1 Set} as the background voices.
The OSI and CSI are enrolled by 5 speakers from the \emph{Test Speaker Set},
forming a speaker group.
The SV is enrolled by
5 speakers from the \emph{Test Speaker Set},
resulting in 5 ivector and 5 GMM systems.

{We conducted the experiments on a server with Ubuntu 16.04 and Intel Xeon CPU E5-2697 v2 2.70GHz with 377G RAM (10 cores). We set $\kappa=0$, max iteration=1,000, max/min learning rate $\eta$
is 1e-3/1e-6, search variance $\sigma$ in NES is 1e-3,
and samples per draw $m$ in NES is $50$, unless explicitly stated.}

%\smallskip
\noindent{\bf {Evaluation metrics}.}
To evaluate our attack,
we use the metrics shown in Table~\ref{tab:metrics}.
{Signal-noise ratio (SNR)} is widely used to quantify the level of signal power to noise power, so we use it to measure the distortion of the adversarial voices~\cite{yuan2018commandersong}. We use the equation, SNR(dB)$ = 10\log_{10}(P_x/P_{\delta})$, to obtain SNR, where $P_x$ is the signal power of the original voice $x$ and $P_{\delta}$ is the power of the perturbation $\delta$. Larger SNR value indicates a (relatively) smaller perturbation.
To evaluate efficiency, we use two metrics: \emph{number of iterations} and \emph{time}.
(Note that the number of queries is the number of iterations multiplied by samples per draw $m$ in NES and $m=50$ in this work.)

  \begin{table}[t]\setlength{\tabcolsep}{4pt}
\caption{{Metrics used in this work}} \label{tab:metrics}
\vspace{-2mm}
\centering
%\scalebox{1}{
\begin{tabular}{l|l}
  \toprule
  {\bf Metric} & {\bf Description} \\\midrule
 \tabincell{l}{ Attack success  rate (ASR)} & \tabincell{l}{Proportion of adversarial voices that\\  are recognized as the target speaker}  \\\midrule
  \tabincell{l}{Untargeted success  rate \\ (UTR) for CSI} & \tabincell{l}{Proportion of adversarial samples that \\are not recognized as the source speaker} \\\midrule
  \tabincell{l}{Untargeted  success  rate \\ (UTR) for OSI} & \tabincell{l}{Proportion of adversarial samples that \\ are not rejected by the target system}
  \\ \bottomrule
\end{tabular}%}
\end{table}

%%%%%%%%%% re-test OSI FAR using different imposter set %%%%%
\begin{figure*}
	\begin{minipage}{0.28\linewidth}
		\tabcaption{Six trained SRSs}
		\vspace{-2mm}
		\label{tab:baseline}
		\scalebox{0.8}{
			\begin{tabular}{c|ccc}
				\toprule
				\textbf{} \textbf{Task} &   \textbf{Metrics} & \textbf{ivector} & \textbf{GMM} \\ \midrule
				{\bf CSI}           & Accuracy & 99.6\%                & 99.3\%           \\ \midrule
				\multirow{2}{*}{\textbf{SV}}     & FRR      & 1.0\%                   & 5.0\%              \\ \cmidrule{2-4}
				& FAR      & 11.0\%                  & 10.4\%           \\ \midrule
				\multirow{3}{*}{\bf OSI} & FRR      & 1.0\%                   & 4.2\%           \\ \cmidrule{2-4}
				& FAR      & 7.9\%                  & 11.2\%           \\ \cmidrule{2-4}
				& OSIER    & 0.2\%                 & 2.8\%           \\ \bottomrule
		\end{tabular} }
	\end{minipage}
%\begin{minipage}{0.28\linewidth}
%	\tabcaption{Six trained SRSs}
%	\label{tab:baseline}
%	\scalebox{0.85}{
%		\begin{tabular}{c|ccc}
%			\toprule
%			\textbf{} \textbf{Task} &   \textbf{Metrics} & \textbf{ivector} & \textbf{GMM} \\ \midrule
%			 {\bf CSI}           & Accuracy & 99.6\%                & 99.3\%           \\ \midrule
%			\multirow{2}{*}{\textbf{SV}}     & FRR      & 1.0\%                   & 5.0\%              \\ \cmidrule{2-4}
%			& FAR      & 11.0\%                  & 10.4\%           \\ \midrule
%			\multirow{3}{*}{\bf OSI} & FRR      & 1.0\%                   & 4.2\%           \\ \cmidrule{2-4}
%			& FAR      & 14.8\%                  & 15.7\%           \\ \cmidrule{2-4}
%			& OSIER    & 0.2\%                 & 2.8\%           \\ \bottomrule
%		\end{tabular} }
% \end{minipage}
 \hspace*{0.1cm}
  \begin{minipage}{0.34\linewidth}
  \centering
	\tabcaption{Results of threshold estimation}
	\vspace{-2mm}
	\label{tab:thresh_estimate_result}
\scalebox{0.8}{\begin{tabular}{ccc|ccc}
		\toprule
		\multicolumn{3}{c}{\bf ivector}                                                   & \multicolumn{3}{c}{\bf GMM}                                                               \\ \midrule
		$\theta$                    & $\acute{\theta}$            & \textbf{Time (s)}                    & $\theta$                     & $\acute{\theta}$              & \textbf{Time (s)}                        \\ \midrule
		\textbf{1.45}		& \textbf{1.47}		&  \textbf{628}		&  \textbf{0.091}		&  \textbf{0.0936}		&  \textbf{157} \\ \midrule
		1.57                        & 1.60                        & 671                   & 0.094                        & 0.0957                        & 260                      \\ \midrule
		1.62                        & 1.64                       & 686                        & 0.106                        & 0.1072                        & 269                        \\ \midrule
		1.73                        & 1.75                       & 750                       & 0.113                        & 0.1141                        & 289                        \\ \midrule
		1.84                        & 1.87                       &  804                      & 0.119                        & 0.1193                        & 314                        \\ \bottomrule
	\end{tabular}}
 \end{minipage}
 \hfill
 \begin{minipage}{0.33\linewidth}
	\centering
	\includegraphics[width=0.85\textwidth]{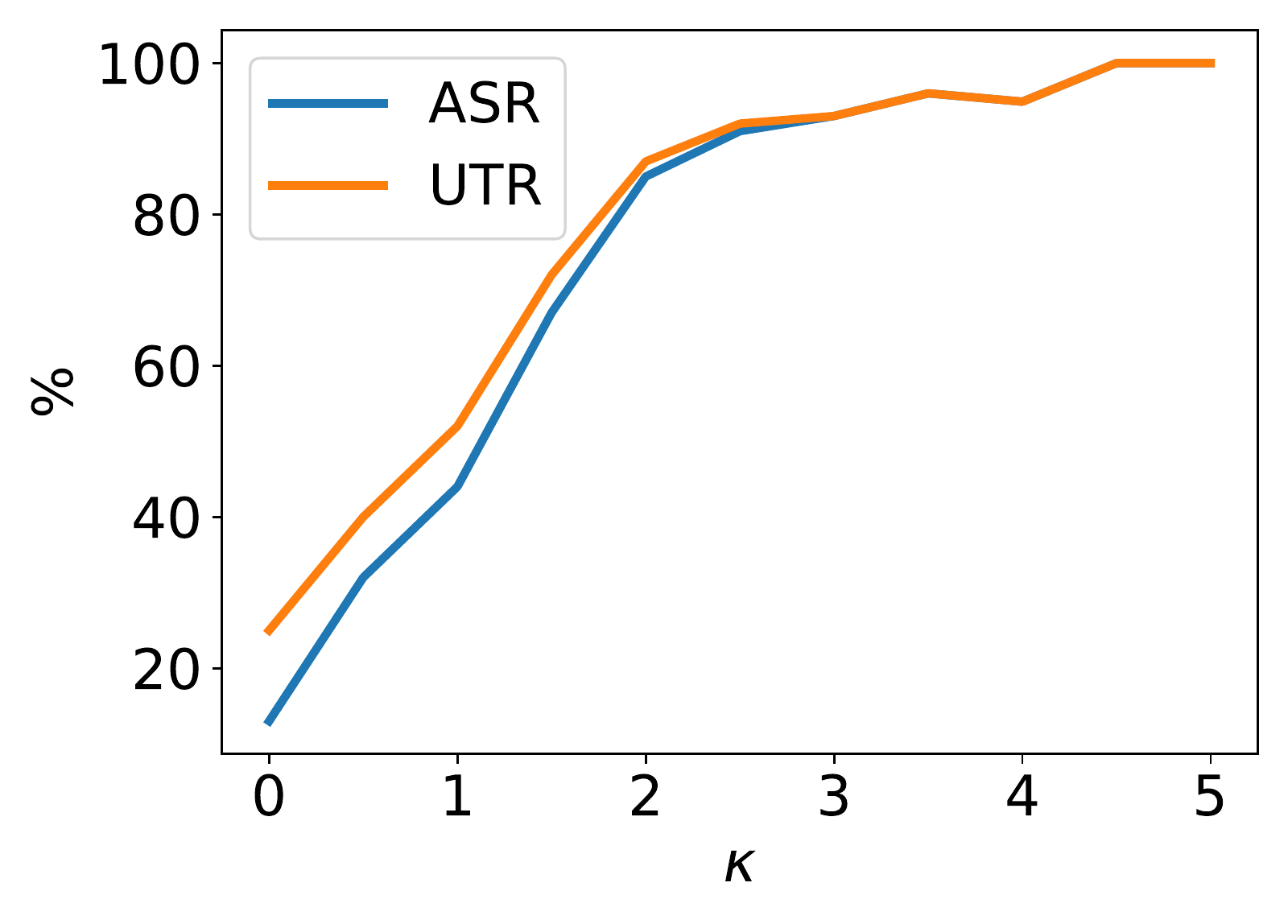}
	\vspace{-2mm}
	\caption{{Transferability rate vs. $\kappa$}}
	\vspace{-1mm}
	\label{fig:tranfer-iteration-versus-kappa}
\end{minipage}
\end{figure*}

%%%%%%%%%%%%%%%%%%%%%%%%%%%%%%%%%%%%%%%%%%%%%%%%%%%%%%%%%%%%%%%%%

%\smallskip
\noindent{\bf {Experiment design}.}
We design five experiments.
(1) We evaluate the \emph{effectiveness} and \emph{efficiency} on both open-source systems {(i.e., ivector, GMM, and xvector)} and
the commercial system Talentedsoft.
We also evaluate \attackname under intra-gender and inter-gender scenarios, as inter-gender attacks are usually more difficult.
(2) We evaluate the \emph{transferability} by attacking the open-source systems with different architecture, training dataset, and parameters,
as well as the commercial system Microsoft Azure.
(3) We further evaluate the \emph{practicability} by playing the adversarial voices over the air in the physical world. (4) For \emph{human-imperceptibility}, we conduct a real human study through Amazon Mechanical Turk platform (MTurk)~\cite{amazon_mturk}, a crowdsourcing marketplace for human intelligence.
(5) We finally evaluate \emph{defense methods}, local smoothing, quantization, audio squeezing, temporal dependency-based detection, to defend against \attackname.

{Recall that we demonstrate our attack on 16 representative attack scenarios out of 40 (cf. \S\ref{subsec:threat}).
In particular, we mainly consider targeted attack which is much more powerful and challenging than untargeted attack~\cite{carlini2018audio}.}
Our experiments suffice to understand the other four parameters of the attack model, i.e.,
inter-gender vs. intra-gender, API vs. over-the-air, OSI vs. CSI vs. SV,
decision and scores vs. decision-only.

The OSI task can be seen as a combination of the CSI and SV tasks  (cf. \S\ref{sec:background}).
Thus, we sometimes only report and analyze the results on the OSI task due to space limitation,
which is much more challenging and representative than the other two.
The missing results can be found in Appendix.

%%%%%%%%%%%%%%%%%%%%%%%%%%%%%%%%%%%%%%%%%%%%%%%%%%%%%%%%%%%%%%%%%

\subsection{{Effectiveness and Efficiency}}\label{sec:c-s-i-experiment}
\noindent{\bf Target model training.}
To evaluate the effectiveness and efficiency,
we train ivector and GMM systems for the OSI, CSI and SV tasks.
The performance of these systems is shown in Table~\ref{tab:baseline},
where accuracy is as usual,
False Acceptance Rate (FAR) is the proportion of voices that are uttered by imposters but accepted by the system~\cite{Homayoon11},
False Rejection Rate (FRR) is the proportion of voices that are uttered by an enrolled speaker but rejected by the system~\cite{Homayoon11}, Open-set Identification Error Rate (OSIER) is the rate of voices
that cannot be correctly classified~\cite{fortuna2005open}.
Notice that the threshold $\theta$ is $1.45$ for ivector and $0.091$ for GMM,
so that the FAR is close to 10\%.
{Although the parameter $\theta$ in SV and OSI tasks can be tuned using Equal Error Rate, i.e., FAR is equal to FRR, we found that the results for SV and OSI tasks do not vary too much (cf. Table~\ref{tab:attack-eer-result} in Appendix).}

%\smallskip
\noindent{\bf Setting.}
The parameter $\epsilon$ is one of the most critical parameters of our attack.
To fine-tune $\epsilon$, we study
ASR, efficiency and distortion by varying $\epsilon$ from 0.05, 0.01, 0.005, 0.004, 0.003, 0.002, to 0.001, on ivector and GMM for the CSI task.
The results are given in Appendix~\ref{sec:tuning}.
With {decreasing} of $\epsilon$, both the attack cost and SNR increase,
while ASR decreases. As a trade-off between ASR, efficiency, and distortion,
we set $\epsilon=0.002$ in this experiment.

The target speakers are the speakers from the Test Speaker Set (cf. Table~\ref{tab:dataset-info}),
the source speakers are the speakers, from the Test Speaker Set for CSI,
and from the Imposter Speaker Set (cf. Table~\ref{tab:dataset-info}) for SV and OSI.
Ideally, we will craft 100 adversarial samples using \attackname for each task,
where 40 adversarial samples are intra-gender and 60 inter-gender for CSI, and 50 intra-gender and 50 inter-gender for SV and OSI.
Note that to diversify experiments, the source speakers of CSI and SV/OSI are
designated to be different.

\smallskip
\noindent{\bf Results.}
The results are shown in Table~\ref{tab:attack_result}.
Since the OSI task is more challenging and representative than the other two, we only analyze the results of the OSI task here.
We can observe that \attackname achieves 99.0\% ASR for both ivector and GMM.
In terms of SNR, the average SNR value is 31.5 (dB) for ivector and 31.4 (dB) for GMM, indicating that the perturbation is less than 0.071\% and 0.072\%.
Furthermore, the average numbers of iterations and execution time are 86 and 38.0 minutes on ivector.
The average numbers of iterations and execution time are 38 and 3.8 minutes on GMM, much smaller than that of ivector.
{Due to space limitation, results of attacking xvector are given in Appendix~\ref{sec:attackDNN} where
we observe similar results.}
These results demonstrate the effectiveness and efficiency of \attackname.

We can also observe that inter-gender attack is much more difficult (more iterations and execution time) than intra-gender attack due to the difference between sounds of male and female.
Moreover, ASR of inter-gender attack is also lower than that of intra-gender attack. The result unveils that once the gender of the target speaker is known by attackers,
it is much easier to launch an intra-gender attack.

For evaluation of the threshold estimation algorithm,
we report the estimated threshold $\acute{\theta}$ in Table~\ref{tab:thresh_estimate_result}  by setting 5 different thresholds.
The estimation error is less than 0.03 for ivector and less than 0.003 for GMM.
This shows that our algorithm is able to effectively estimate the threshold in less than 13.4 minutes.
Note that our attack is black-box, and the actual thresholds are accessed  \emph{only} for
evaluation.

\begin{table*}
	\caption{Experimental results of \attackname when $\epsilon=0.002$, where \#Iter refers to \#Iteration.}
	\vspace{-1mm}
	\label{tab:attack_result}
	\centering \setlength{\tabcolsep}{2pt}
	\scalebox{0.95}{%
		\begin{tabular}{c|cccc|cccc|cccc|cccc|cccc|cccc}
			\toprule
			\multirow{4}{*}{\textbf{Task}} & \multicolumn{8}{c|}{\textbf{System}} & \multicolumn{8}{c|}{\textbf{System (Intra-gender attack)}} & \multicolumn{8}{c}{\textbf{System (Inter-gender attack)}} \\ \cmidrule{2-25}
			& \multicolumn{4}{c|}{\textbf{ivector}} & \multicolumn{4}{c|}{\textbf{GMM}} & \multicolumn{4}{c|}{\textbf{ivector}} & \multicolumn{4}{c|}{\textbf{GMM}} & \multicolumn{4}{c|}{\textbf{ivector}} & \multicolumn{4}{c}{\textbf{GMM}} \\ \cmidrule{2-25}
			& \textbf{\#Iter} & \textbf{\begin{tabular}[c]{@{}c@{}}Time\\ (s)\end{tabular}} & \textbf{\begin{tabular}[c]{@{}c@{}}SNR\\ (dB)\end{tabular}} & \textbf{\begin{tabular}[c]{@{}c@{}}ASR\\ (\%)\end{tabular}} & \textbf{\#Iter} & \textbf{\begin{tabular}[c]{@{}c@{}}Time\\ (s)\end{tabular}} & \textbf{\begin{tabular}[c]{@{}c@{}}SNR\\ (dB)\end{tabular}} & \textbf{\begin{tabular}[c]{@{}c@{}}ASR\\ (\%)\end{tabular}} & \textbf{\#Iter} & \textbf{\begin{tabular}[c]{@{}c@{}}Time\\ (s)\end{tabular}} & \textbf{\begin{tabular}[c]{@{}c@{}}SNR\\ (dB)\end{tabular}} & \textbf{\begin{tabular}[c]{@{}c@{}}ASR\\ (\%)\end{tabular}} & \textbf{\#Iter} & \textbf{\begin{tabular}[c]{@{}c@{}}Tine\\ (s)\end{tabular}} & \textbf{\begin{tabular}[c]{@{}c@{}}SNR\\ (dB)\end{tabular}} & \textbf{\begin{tabular}[c]{@{}c@{}}ASR\\ (\%)\end{tabular}} & \textbf{\#Iter} & \textbf{\begin{tabular}[c]{@{}c@{}}Time\\ (s)\end{tabular}} & \textbf{\begin{tabular}[c]{@{}c@{}}SNR\\ (dB)\end{tabular}} & \textbf{\begin{tabular}[c]{@{}c@{}}ASR\\ (\%)\end{tabular}} & \textbf{\#Iter} & \textbf{\begin{tabular}[c]{@{}c@{}}Time\\ (s)\end{tabular}} & \textbf{\begin{tabular}[c]{@{}c@{}}SNR\\ (dB)\end{tabular}} & \textbf{\begin{tabular}[c]{@{}c@{}}ASR\\ (\%)\end{tabular}} \\ \midrule
			{\bf CSI} & 124 & 2845 & 30.2 & 99.0 & 40 & 218 & 29.3 & 99.0 & 92 & 2115 & 29.3 & 100.0 & 25 & 126 & 28.8 & 100.0 & 146 & 3340 & 30.8 & 98.0 & 50 & 278 & 29.62 & 98.0 \\ \midrule
			{\bf SV} & 84 & 2014 & 31.6 & 99.0 & 39 & 241 & 31.4 & 99.0 & 31 & 751 & 31.7 & 98.0 & 30 & 185 & 31.7 & 100.0 & 135 & 3252 & 31.6 & 100.0 & 48 & 298 & 31.2 & 98.0 \\ \midrule
			{\bf OSI} & 86 & 2277 & 31.5 & 99.0 & 38 & 226 & 31.4 & 99.0 & 32 & 833 & 31.3 & 98.0 & 31 & 178 & 31.5 & 100.0 & 140 & 3692 & 31.6 & 100.0 & 45 & 274 & 31.2 & 98.0 \\ \bottomrule
		\end{tabular}%
	}
\end{table*}

%%%%%%%%%%%%%%%%%%%%%%%%%%%%%%%%%%%%%%%%%%%%%%%%%%%%%%%%%%%%%%%%%

%smallskip
\noindent \textbf{Attacking the commercial system Talentedsoft~\cite{Talentedsoft}.}
We also evaluate the effectiveness and efficiency of \attackname on Talentedsoft,
developed by the constitutor of the voiceprint recognition industry standard of the Ministry of Public Security (China).
We query this online platform via the HTTP post (seen as the exposed API).
Since Talentedsoft targets Chinese Mandarin,
to fairly test Talentedsoft, we use the Chinese Mandarin voice database aishell-1~\cite{aishell-1}.
Both FAR and FRR of Talentedsoft are 0.15\%, tested using 20 speakers and 7,176 voices in total which are randomly chosen from aishell-1.

{We enroll 5 randomly chosen speakers from aishell-1 as targeted speakers,
resulting in 5 SV systems.
Each of them is attacked using another 20 randomly chosen speakers and one randomly chosen voice per speaker.
Our attack achieves 100\% ASR within 50 iterations (i.e., 2,500 queries) on average.
Remark that \attackname is an iterative-based method. We can always set some time slot between iterations or queries so that such amount of queries do not cause heavy traffic burden to the server, hence our attack is feasible.}
This demonstrates the effectiveness and efficiency of \attackname on commercial systems that are completely black-box.

%%%%%%%%%%%%%%%%%%%%%%%%%%%%%%%%%%%%%%%%%%%%%%%%%%%%%%%%%%%%%%%%%

\subsection{Transferability}\label{sec:transferbility}
%\highlight{[Reviewer1: deploy on DNN models. Reviewer4: 2dB, 0dB corresponds to the signal having equal loudness to the added noise.]}
Transferability~\cite{szegedy2013intriguing}
is the property that some
adversarial samples produced to mislead a model (called source system)
can mislead other models (called target system) even if their architectures, training datasets,
or parameters differ.

\begin{figure}[t]
	\centering
	\includegraphics[width=0.35\textwidth]{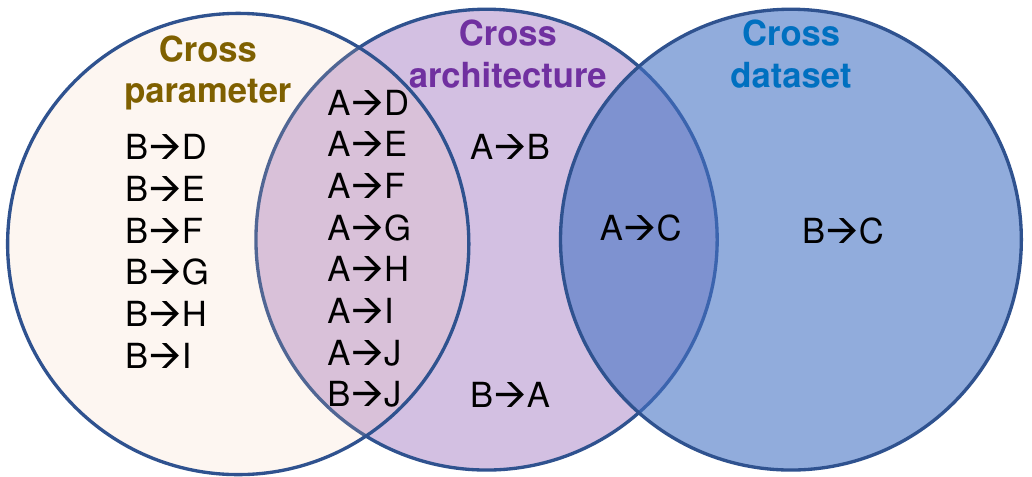}
	\vspace{-1mm}
	\captionof{figure}{{Distribution of transferability attacks}}
	\label{fig:transferbility-type}
\end{figure}

\smallskip
\noindent{\bf Setting.}
To evaluate the transferability, % of the adversarial voices crafted by \attackname,
we regard the previously built GMM (A) and ivector (B) as source systems
{ and build
another 8 target systems (denoted by C,\dots,J respectively).
C,\dots,I are ivector systems differing in key parameter and training dataset, and J is the xvector system.
For details and performance of these systems, refer to Tables~\ref{tab:system_info} and~\ref{tab:eight-systems-performance} in Appendix.
We denote by $X\to Y$ the transferability attack
where $X$ is the source system and $Y$ is the target system.}
The distribution of the transferability attacks is shown in Fig.~\ref{fig:transferbility-type} in terms
of architecture, training dataset, and key parameters.
We can see that some attacks belong to multiple scenarios.
{We set $\epsilon=0.05$ and
(1) $\kappa=0.2$ (GMM) and $\kappa=10$ (ivector) for the CSI task,
(2) $\kappa=3$ (GMM) and $\kappa=4$ (ivector) for the SV task,
(3) $\kappa=3$ (GMM) and $\kappa=5$ (ivector) for the OSI task.
Remark that $\kappa$ differs from architectures and tasks
due to their different scoring mechanisms. We fine-tuned
the parameter $\kappa$ for ASR under the max iteration bound 1,000.}

%%%%%%%%%%%%%%%%%%%%%%%%%%%%%%%%%%%%%%%%%%%%%%%%%%%%%%%%%%%%%%%%%

\begin{table*}\renewcommand{\arraystretch}{1.3}
\centering\setlength{\tabcolsep}{3.5pt}
\caption{{Transferability rate (\%) for OSI task, where S and T denote source and target systems respectively.}}
\vspace{-1mm}
\label{tab:best-transferability-result}
%\resizebox{\textwidth}{!}{%
\scalebox{0.95}{  \begin{tabular}{c|c|c|c|c|c|c|c|c|c|c|c|c|c|c|c|c|c|c|c|c}
\hline
\multirow{2}{*}{\textbf{\diagbox[]{S}{T}}} & \multicolumn{2}{c|}{\textbf{A}} & \multicolumn{2}{c|}{\textbf{B}} & \multicolumn{2}{c|}{\textbf{C}} & \multicolumn{2}{c|}{\textbf{D}} & \multicolumn{2}{c|}{\textbf{E}} & \multicolumn{2}{c|}{\textbf{F}} & \multicolumn{2}{c|}{\textbf{G}} & \multicolumn{2}{c|}{\textbf{H}} & \multicolumn{2}{c|}{\textbf{I}} & \multicolumn{2}{c}{\textbf{J}} \\ \cline{2-21}
 & \textbf{ASR} & \textbf{UTR} & \textbf{ASR} & \textbf{UTR} & \textbf{ASR} & \textbf{UTR} & \textbf{ASR} & \textbf{UTR} & \textbf{ATR} & \textbf{UTR} & \textbf{ASR} & \textbf{UTR} & \textbf{ASR} & \textbf{UTR} & \textbf{ASR} & \textbf{UTR} & \textbf{ASR} & \textbf{UTR} & \textbf{ASR} & \textbf{UTR} \\ \hline
\textbf{A} & --- & --- & 62.0 & 64.0 & 48.0 & 48.0 & 55.2 & 56.9 & {68.0} & {68.0} & 64.0 & 64.0 & 52.0 & 54.0 & {68.0} & {68.0} & 38.0 & 40.0 & 34.0 & 42.0 \\ \hline
\textbf{B} & 5.0 & 5.0 & --- & --- & 67.5 & 67.5 & \textbf{100.0} & \textbf{100.0} & \textbf{100.0} & \textbf{100.0} & \textbf{100.0} & \textbf{100.0} & \textbf{100.0} & \textbf{100.0} & \textbf{100.0} & \textbf{100.0} & 72.5 & 75.0 & 40.0 & 41.7 \\ \hline
\end{tabular}}%
\end{table*}

%\smallskip
\noindent{\bf Results.}
{The results of attacking OSI systems
are shown in Table~\ref{tab:best-transferability-result}.
All the attacks (except for $B\to A$) achieve 34\%-68\% ASR and 40\%-100\% UTR.
For $B\to D$, $B\to E$, $B\to F$, $B\to G$, and $B\to H$ (all are ivector, but differ in one key parameter),
\attackname achieves 100\% ASR and UTR, indicating that cross architecture reduces transferability rate.
From $A\to B$ and $A\to C$ (where $A$ is GMM, $B$ and $C$ are ivector but differ in training data),
cross dataset also reduces transferability rate.
The transferability rate of $B\to A$ is the lowest one and less than that of $A\to B$,
indicating that transferring from the architecture ivector ($B$)
to GMM ($A$) is more difficult.
Compared with $A\to C$ (both cross dataset and architecture),
$B \to C$ (cross dataset) achieves nearly 20\% more ASR and UTR.
This reveals that the larger the difference between the source and target systems is,
the more difficult the transferability attack is.
Due to space limitation, the results of attacking the CSI and SV systems are
shown in Tables~\ref{tab:best-transferability-result-csi} and~\ref{tab:best-transferability-result-sv} in Appendix.
We can observe similar results.
The average SNR is similar to the one given in Table~\ref{tab:over-the-air-basic-result}.}

{To understand how the value of $\kappa$ influences the transferability rate,
we conduct $B\to F$ attack (OSI task) by fixing $\epsilon=0.05$ and varying $\kappa$ from 0.5 to 5.0 with step 0.5.
In this experiment, the number of iterations is unlimited.
The results are shown in Fig.~\ref{fig:tranfer-iteration-versus-kappa}.
Both ASR and UTR increase quickly with $\kappa$,
and reach 100\% when $\kappa= 4.5$.
This demonstrates that increasing the value of $\kappa$ increases the probability of a successful transferability attack.}
%It is unsurprising that attack cost also increases with $\kappa$, but increases relatively slowly
%when $\kappa<4$.

%%%%%%%%%%%%%%%%%%%%%%%%%%%%%%%%%%%%%%%%%%%%%%%%%%%%%%%%%%%%%%%%%

%\smallskip
\noindent{\bf Attacking the commercial system Microsoft Azure~\cite{Azure}.}
Microsoft Azure is a {cloud service platform} with the second largest market share in the world.
It supports both the SV and OSI tasks via HTTP REST API.
Unlike Talentedsoft, Azure's API only returns the decision (i.e., the predicted speaker) along with 3 confidence levels (i.e., low, normal and high) instead of scores,
so we attack this platform via transferability.
We enroll 5 speakers from the Test Speaker Set to build an OSI system on Azure {(called OSI-Azure for simplicity)}.  {Its FAR is $0\%$ tested by the Imposter Speaker Set.
For each target speaker,
we randomly select 10 source speakers and 2 voices per source speaker from LibriSpeech,
which are rejected by OSI-Azure.
We set $\epsilon=0.05$ and craft 100 adversarial voices on the GMM system, as it produces high tranferability rate in the above experiment.
The ASR, UTR and SNR are 26.0\%, 41.0\% and 6.8 dB, respectively.
They become 34.0\%, 57.0\% and 2.2 dB when we increase $\epsilon$ from $0.05$ to $0.1$.}

%%%%%%%%%%% Azure SV Revise %%%%%%%%%%%%%%%%%%%%%%%%%%%%

{{We also demonstrate \attackname on the SV task of Azure (SV-Azure) which
is text-dependent with 10 supported texts.
% dynamically and randomly generates text from 10 fixed texts.
We recruited and asked 2 speakers to read each text 10 times, resulting in 200 voices.
%Note that we do not exploit Text-to-Speech Engine~\cite{cereproc_tts} directly because the generated voices are in lower quality than genuine human voices.
%Since SV-Azure is text-dependent,
For each pair of speaker and text,
we randomly select 3 enrollment voices for both GMM and SV-Azure, and the FARs of them are 0\%.
% we randomly select 3 enrollment voices for both GMM and SV-Azure,
% and test the performance using the other 97 voices.
% we enroll three voices into GMM and Azure
% since SV-Azure requires three recordings for one enrollment.
% The baseline of SV-Azure (resp. GMM SV) is FAR=$0\%$ and FRR=$1.4\%$ (resp. FAR=$0\%$ and FRR=$4.3\%$).
% The FARs of them are 0\%. % tested by the other 97 voices.
% \chengk{Since SV-Azure is text-dependent,
% for each pair of speaker and text,
% we first need to construct a SV system on both GMM and Azure.
% The baseline of SV-Azure (resp. GMM SV) is FAR=$0\%$ and FRR=$1.4\%$ (resp. FAR=$0\%$ and FRR=$4.3\%$). @guangke, revise it.}
% We set $\epsilon=0.05$ and craft 200 adversarial samples on the GMM system.
%After that, we conduct a transferability attack on SV-Azure.
We attack SV-Azure using 200 adversarial samples crafted from GMM ($\epsilon=0.05$, $\kappa=3$).
However, SV-Azure reports ``error, too noisy" instead of ``accept" or ``reject" for 190 adversarial voices.
Among the other 10 voices, one voice is accepted, leading to 10\% ASR.
To our knowledge, this is the first time that SV-Azure is successfully attacked.
As Azure is proprietary without any publicly available information,
it is very difficult to know the reason why SV-Azure outputs ``error, too noisy".
After comparing the SNR of the 190 voices with
the other 10 voices (8.8 dB vs. 11.5 dB),
we suspect that it checks each input
and outputs ``error, too noisy" without model classification if the noise of the input is too large.
This check makes SV-Azure more challenging to attack,
% but we guess
but we infer
it may also reject normal voices when the background is noisy in practice.
%which is text-independent.
}
}
% \fu{To be resived!!!}

\subsection{{Practicability for Over-the-Air Attack}}\label{subsec:over}
{To simulate over-the-air attack in the physical world,
we first craft adversarial samples by directly interacting with API of the system (i.e., over the line),
then play and record these adversarial voices via loudspeakers and microphones,
and finally send recorded voices to the system via API to check their effectiveness.
% and finally send recorded voices to the target system via API to check their effectiveness.
Our experiments are conducted in an indoor room (length, width, and height are 10, 4, 3.5 meters).
% Our experiments are conducted in a meeting room (length, width, and height are 12, 8, 4 meters).
To thoroughly evaluate \attackname, the over-the-air attacks
vary in systems, devices (loudspeakers and microphones), distance between loudspeakers and microphones, and acoustic environments. In total, it covers 26 scenarios.
The overview of different settings is shown in Table~\ref{tab:over-the-air-four-expers} in Appendix.
% For instances, we consider all tasks of ivector and GMM, and the OSI-Azure only,
% We consider all tasks of ivector and GMM, and the OSI-Azure only,
% % as SV-Azure is impractical for normal voices (cf. Section~\ref{sec:transferbility}).
% \chengk{as only one adversarial voice can be transferred to SV-Azure (cf. Section~\ref{sec:transferbility}).}
We consider all tasks of ivector and GMM, and the OSI-Azure only.
% For open-source systems, the source and target systems are identical.
We use the same parameters as in Section~\ref{sec:transferbility},
as over-the-air attack is more practical yet more challenging due to
the noise introduced from both air channel and electronic devices
which probably disrupts the perturbations of adversarial samples.
For OSI-Azure, we use the adversarial voices crafted on GMM in Section~\ref{sec:transferbility} that are successfully transferred to OSI-Azure.}
% \chengk{Notice: If SV-Azure transferability attack is successful, the above sentence "as SV-Azure is impractical..." needs to be modified and additional over-the-air experiment is needed.}

%For each pair of source and target systems,

\begin{table}\renewcommand{\arraystretch}{1.3}
\centering
% \caption{{Results of different target systems}}
\caption{{Results of different systems}}
\vspace{-1mm}
\setlength{\tabcolsep}{2pt}
\label{tab:over-the-air-basic-result}
\resizebox{0.48\textwidth}{!}{%
 \begin{tabular}{c|c|c|c|c}
\hline
\multicolumn{2}{c|}{\multirow{2}{*}{\textbf{System}}} & \multirow{2}{*}{\textbf{\begin{tabular}[c]{@{}c@{}}SNR\\ (dB)\end{tabular}}} & \multicolumn{2}{c}{\textbf{Result (\%)}} \\ \cline{4-5}
\multicolumn{2}{c|}{} &  & \textbf{\begin{tabular}[c]{@{}c@{}}Normal voices \end{tabular}} & \textbf{\begin{tabular}[c]{@{}c@{}}Adversarial voices \end{tabular}} \\ \hline
\multirow{3}{*}{\textbf{ivector}} & \textbf{CSI} & 6.6 & Accuracy: 100 & ASR: 80, UTR: 80 \\ \cline{2-5}
 & \textbf{SV} & 9.8 & FAR: 0, FRR: 0 & ASR: 76 \\ \cline{2-5}
 & \textbf{OSI} & 7.8 & FAR: 4, FRR: 0, OSIER: 0 & ASR: 100, UTR: 100 \\ \hline
\multirow{3}{*}{\textbf{GMM}} & \textbf{CSI} & 6.1 & Accuracy: 85 & ASR: 90, UTR: 100 \\ \cline{2-5}
 & \textbf{SV} & 7.9 & FAR: 0, FRR: 62 & ASR: 100 \\ \cline{2-5}
 & \textbf{OSI} & 8.2 & FAR: 0, FRR: 65, OSIER: 0 & ASR: 100, UTR: 100 \\ \hline
\textbf{Azure} & \textbf{OSI} & 6.8 & FAR: 5, FRR: 2, OSIER: 0 & ASR: 70, UTR: 70\\ \hline
\end{tabular}%
}
\end{table}

\smallskip
\noindent
\textbf{Results of different systems.}
We use portable speaker (JBL clip3~\cite{JBL-clip3}) as
the loudspeaker, iPhone 6 Plus (iOS) as the microphone with
1 meter distance between them. We attack all tasks of ivector and GMM, and the OSI-Azure in a relatively quiet environment. The results are shown in Table~\ref{tab:over-the-air-basic-result}.
We can observe that the FRR of GMM SV (resp. OSI) is 62\% (resp. 65\%),
revealing that GMM is less robust than ivector
for normal voices.
{\attackname achieves
(1) for the CSI task, 90\% ASR (i.e., the system classifies the adversarial voice as the target speaker) and 100\% UTR
(i.e., the system does not classify the adversarial voice as the source speaker) on the GMM,
and achieves 80\% ASR and 80\% UTR on the ivector;
(2) for the SV task, at least 76\% ASR;
(3) for the OSI task, 100\% ASR on both the GMM and ivector;
(4) achieves 70\% ASR on the commercial system OSI-Azure.}

{In terms of SNR, the average SNR is no less than 6.1 dB, and the average SNR is up to 9.8 dB on the ivector for the SV task, indicating that the power of the signal is 9.5 times greater than that of the noise.
Moreover, the SNR is much better than the over-the-air attack in CommanderSong~\cite{yuan2018commandersong}.
% This demonstrates the effectiveness of \attackname against different
% target systems when playing over the air in the physical world.
% This demonstrates the effectiveness of \attackname against different systems when playing over the air in the physical world.
}

\smallskip
\noindent
{\textbf{Results of different devices.}
For loudspeakers, we use 3 common devices: laptop (DELL), portable speaker (JBL clip3)
and broadcast equipment (Shinco~\cite{Shinco-broadcast}).
For microphones, we use built-in microphones of 2 mobile phones: OPPO (Android) and iPhone 6 Plus (iOS).
We evaluate \attackname against the OSI task of
ivector with 1 meter distance in a relatively quiet environment.
The results are shown in Table~\ref{tab:over-the-air-hardware-result}.}

{We can observe that
for any pair of loudspeaker and microphone,
\attackname can achieve at least 75\% ASR and UTR.
When JBL clip3 or DELL is the loudspeaker
and iPhone 6 Plus is the microphones, \attackname is able to achieve 100\% ASR.
When the loudspeaker is fixed, the ASR
and UTR of attacks using IPhone 6 Plus are higher (at least 14\% and 16\% more) than
that of using OPPO.
Possible reason is that the sound quality of iPhone 6 Plus is better than that of OPPO phone.
These results demonstrate the effectiveness of \attackname on various devices.}

\smallskip
\noindent
{{\bf Results of different distances.}
% To understand the impact of the distance between loudspeakers and microphones,
% we use JBL clip3  as the loudspeaker and iPhone 6 Plus as the microphone.
% We attack the OSI task of ivector in a relatively
% quiet environment by varying distance from 0.25, 0.5, 1, 2, 4 to 8 meters.
To understand the impact of the distance between loudspeakers and microphones,
we vary distance from 0.25, 0.5, 1, 2, 4 to 8 meters.
We attack the OSI task of ivector in a relatively
quiet environment by using JBL clip3 as the loudspeaker and iPhone 6 Plus as the microphone.}

The results are shown in Table~\ref{tab:over-the-air-distance-result}.
We can observe that \attackname can achieve 100\% ASR and UTR when
the distance is no more than 1 meter.
When the distance is increased to 2 meters (resp. 4 meters),
ASR and UTR drop to 70\% (resp. 40\% and 50\%).
Although ASR and UTR drop to 10\% when the distance is 8 meters,
FRR also increases to 32\%.
This shows the effectiveness of \attackname
under different distances.

\begin{table}[t]\renewcommand{\arraystretch}{1.2}
\centering
\caption{{Results of different devices (\%), where L and M denote loudspeakers and microphones respectively.}}
\vspace{-1mm}
\setlength{\tabcolsep}{1.5pt}
\label{tab:over-the-air-hardware-result}
\resizebox{0.49\textwidth}{!}{%
  \begin{tabular}{c|c|c|c|c|c|c|c|c|c|c}
\hline
\multirow{3}{*}{\textbf{\diagbox[]{L}{M}}} & \multicolumn{5}{c|}{\textbf{iPhone 6 Plus (iOS)}} & \multicolumn{5}{c}{\textbf{OPPO (Android)}} \\ \cline{2-11}
 & \multicolumn{3}{c|}{\textbf{Normal voices}} & \multicolumn{2}{c|}{\textbf{Adv. voices}} & \multicolumn{3}{c|}{\textbf{Normal voices}} & \multicolumn{2}{c}{\textbf{Adv. voices}} \\ \cline{2-11}
 & \textbf{FAR} & \textbf{FRR} & \textbf{OSIER} & \textbf{ASR} & \textbf{UTR} & \textbf{FAR} & \textbf{FRR} & \textbf{OSIER} & \textbf{ASR} & \textbf{UTR} \\ \hline
\textbf{DELL} & 10 & 0 & 0 & 100 & 100 & 13 & 6 & 0 & 78 & 80 \\ \hline
\textbf{JBL clip3} & 4 & 0 & 0 & 100 & 100 & 6 & 0 & 0 & 80 & 80 \\ \hline
\textbf{Shinco} & 8 & 5 & 0 & 89 & 91 & 14 & 0 & 0 & 75 & 75 \\ \hline
\end{tabular}}
\end{table}

\begin{table*}\renewcommand{\arraystretch}{1.2}
\begin{minipage}{0.355\linewidth}%
\centering\setlength{\tabcolsep}{3pt}
\caption{{Results of different distances (\%)}}
%\vspace{-1mm}
\label{tab:over-the-air-distance-result}
\scalebox{0.99}
{ \begin{tabular}{c|c|c|c|c|c|c|c}
\hline
\multicolumn{2}{c|}{{\bf Distance} (meter)}  & \textbf{0.25} & \textbf{0.5} & \textbf{1} & \textbf{2} & \textbf{4} & \textbf{8} \\ \hline
\multirow{3}{*}{\textbf{\begin{tabular}[c]{@{}c@{}}Normal\\ Voices\end{tabular}}} & \textbf{FAR} & 4 & 3 & 4 & 6 & 0 & 0 \\ \cline{2-8}
 & \textbf{FRR} & 0 & 0 & 0 & 5 & 10 & 32 \\ \cline{2-8}
 & \textbf{OSIER} & 0 & 0 & 0 & 0 & 0 & 0 \\ \hline
\multirow{2}{*}{\textbf{\begin{tabular}[c]{@{}c@{}}Adversarial\\ Voices\end{tabular}}} & \textbf{ASR} & 100 & 100 & 100 & 70 & 40 & 10 \\ \cline{2-8}
 & \textbf{UTR} & 100 & 100 & 100 & 70 & 50 & 10 \\ \hline
\end{tabular} }%
\end{minipage}
\hfill
\begin{minipage}{0.64\linewidth}%
\centering\setlength{\tabcolsep}{2pt}
\caption{{Results of different acoustic environments (\%)}}
%\vspace{-1mm}
\label{tab:over-the-air-acoustic-result}
\scalebox{0.9}{ \begin{tabular}{c|c|c|c|c|c|c|c|c|c|c|c}
\hline
\multicolumn{2}{c|}{\textbf{Environment}} & \textbf{Quiet} & \tabincell{c}{\textbf{White} \\ (45 dB)} & \tabincell{c}{\textbf{White} \\ (50 dB)}  & \tabincell{c}{\textbf{White} \\ (60 dB)}  & \tabincell{c}{\textbf{White} \\ (65 dB)}  & \tabincell{c}{\textbf{White} \\ (75 dB)}  & \tabincell{c}{\textbf{Bus} \\ (60 dB)}  &  \tabincell{c}{\textbf{Rest.} \\ (60 dB)}  & \tabincell{c}{\textbf{Music} \\ (60 dB)}  & \tabincell{c}{\textbf{Abs. Music} \\ (60 dB)} \\ \hline
\multirow{3}{*}{\textbf{\begin{tabular}[c]{@{}c@{}}Normal\\ voices\end{tabular}}} & \textbf{FAR} & 4 & 0 & 6 & 0 & 0 & 10 & 0 & 0 & 0 & 4 \\ \cline{2-12}
 & \textbf{FRR} & 0 & 5 & 12 & 30 & \textbf{40} & \textbf{97} & 25 & 20 & 10 & 10 \\ \cline{2-12}
 & \textbf{OSIER} & 0 & 0 & 0 & 0 & 0 & 0 & 0 & 0 & 10 & 0 \\ \hline
\multirow{2}{*}{\textbf{\begin{tabular}[c]{@{}c@{}}Adv.\\ voices\end{tabular}}} & \textbf{ASR} & 100 & 75 & 70 & 57 & 20 & 2 & 50 & 50 & 66 & 48 \\ \cline{2-12}
 & \textbf{UTR} & 100 & 75 & 70 & 60 & 20 & 2 & 50 & 50 & 67 & 48 \\ \hline
\end{tabular}}
\end{minipage}
\end{table*}

\smallskip
\noindent
{{\bf Results of different acoustic environments}.
We attack the OSI task of ivector using JBL clip3 and iPhone 6 Plus with 1 meter distance.
To simulate different acoustic environments,
% we play different types of noises in the background
% using Shinco that is 1 meter away from both the loudspeaker
% and microphone.
we play different types of noises in the background
using Shinco broadcast equipment.
Specifically, we select 5 types of noises from Google AudioSet~\cite{google-audio-set}: white noise, bus noise, restaurant noise, music noise, and absolute music noise. White noise is widespread in nature, while bus, restaurant, (absolute) music noises are representative of several daily life scenarios where \attackname may be launched.
For white noise, we vary its volume from 45 dB to 75 dB, while
the volumes of other noises are 60 dB.
Both adversarial and normal voices are played at 65 dB on average.
The results are shown in Table~\ref{tab:over-the-air-acoustic-result}.}

{We can observe that \attackname achieves at least 48\% ASR and UTR
when the volume of background noises is no more than 60 dB no matter
the type of the noises.
Although both ASR and UTR decrease with increasing
the volume of white noises,
the FRR also increases quickly.
This demonstrates the effectiveness of
\attackname in different acoustic environments.}

%%%%%%%%%%%%%%%%%%%%%%%%%%%%%%%%%%%%%%%%%%%%%%%%%%%%%%%%%%%%%%%%%

\subsection{{Human-Imperceptibility via Human Study}}\label{sec:human-study-exper}
%\sen{[For reviewer1] Provide more details if necessary. Release the samples used in user study? If we do not consider some conditions in our user study, we need to explain, such as language, age.}
To demonstrate the imperceptibility of adversarial samples, we conduct a human study on MTurk~\cite{amazon_mturk}.
The survey is approved by the Institutional Review Board (IRB) of our institutes.

\smallskip
\noindent \textbf{Setup of human study.}
We recruit participants from MTurk and ask them to choose one of the two tasks and finish the corresponding questionnaire.
{We neither reveal the purpose of our study to the participants, nor record personal information of participants such as first language, age and region. %The datasets in Table~\ref{tab:dataset-info} only contain gender of speakers.
The Amazon MTurk  has designed Acceptable Use Policy for permitted and prohibited uses of MTurk, which prohibits bots or scripts or other automated answering tools to complete Human Intelligence Tasks~\cite{AUP19}. Thus, we argue that the number of participants can reasonably guarantee the diversity of participants.}
The two tasks are described as follows.

\begin{itemize}
    \item \textit{Task 1: Clean or Noisy.}
    This task asks participants to tell whether the playing voice is clean or noisy. Specifically, we randomly select $12$ original voices and
    $15$ adversarial voices crafted from other original voices,
    among which 12 adversarial voices are randomly selected from the voices which become non-adversarial (called ineffective) when playing over the air with $\epsilon=0.002$ and low confidence,
    and the other 3 are randomly selected from the voices which remain adversarial (called effective) when playing over the air with $\epsilon=0.1$ and high confidence.
    We ask users to choose whether a voice has any background noise (The three options are \textit{clean}, \textit{noisy}, and \textit{not sure}).

\item \textit{Task 2: Identify the Speaker.} This task asks participants to tell whether the voices in a pair are uttered by the same speaker.
Specifically, we randomly select 3 speakers (2 male and 1 female), and randomly choose 1 normal voice per speaker (called reference voice).
Then for each speaker, we randomly select $3$ normal voices, $3$ distinct adversarial voices that are crafted from other normal voices of the same speaker,
and $3$ normal voices from other speakers. %, resulting in 9 pairs of voices along per reference voice.
In summary, we build 27 pairs of voices:
9 pairs are \textit{normal pairs} (one reference voice and one normal voice from the same speaker),
9 pairs are \textit{other pairs} (one reference voice and one normal voice from another speaker) and
9 pairs are \textit{adversarial pairs} (one reference voice and one adversarial voice from the same speaker).
{Among 9 adversarial pairs, 6 pairs contain effective adversarial samples when playing over the air, and 3 pairs do not.}
We ask the participants to tell whether the voices in each pair are uttered by the same speaker (The three options are  \textit{same}, \textit{different}, and \textit{not sure}).
\end{itemize}

{To ensure the quality of our questionnaire and validity of our results, we filter out the questionnaires that are randomly chosen by participants. In particular, we set three simple questions in each task. For task 1, we insert three silent voices as a concentration test. For task 2,  we insert three pairs of voices, where each pair contains one male voice and one female voice as a concentration test. Only when all of them are correctly answered, we regard it as a valid questionnaire, otherwise, we exclude it.}

%%%%%%%%%%%%%%%%%%%%%%%%%%%%%%%%%%%%%%%%%%%%%%%%%%%%%%%%%%%%%%%%%

\begin{figure}[t]
	\centering
	\begin{subfigure}[t]{0.25\textwidth}
		\centering
		\includegraphics[width=1\textwidth]{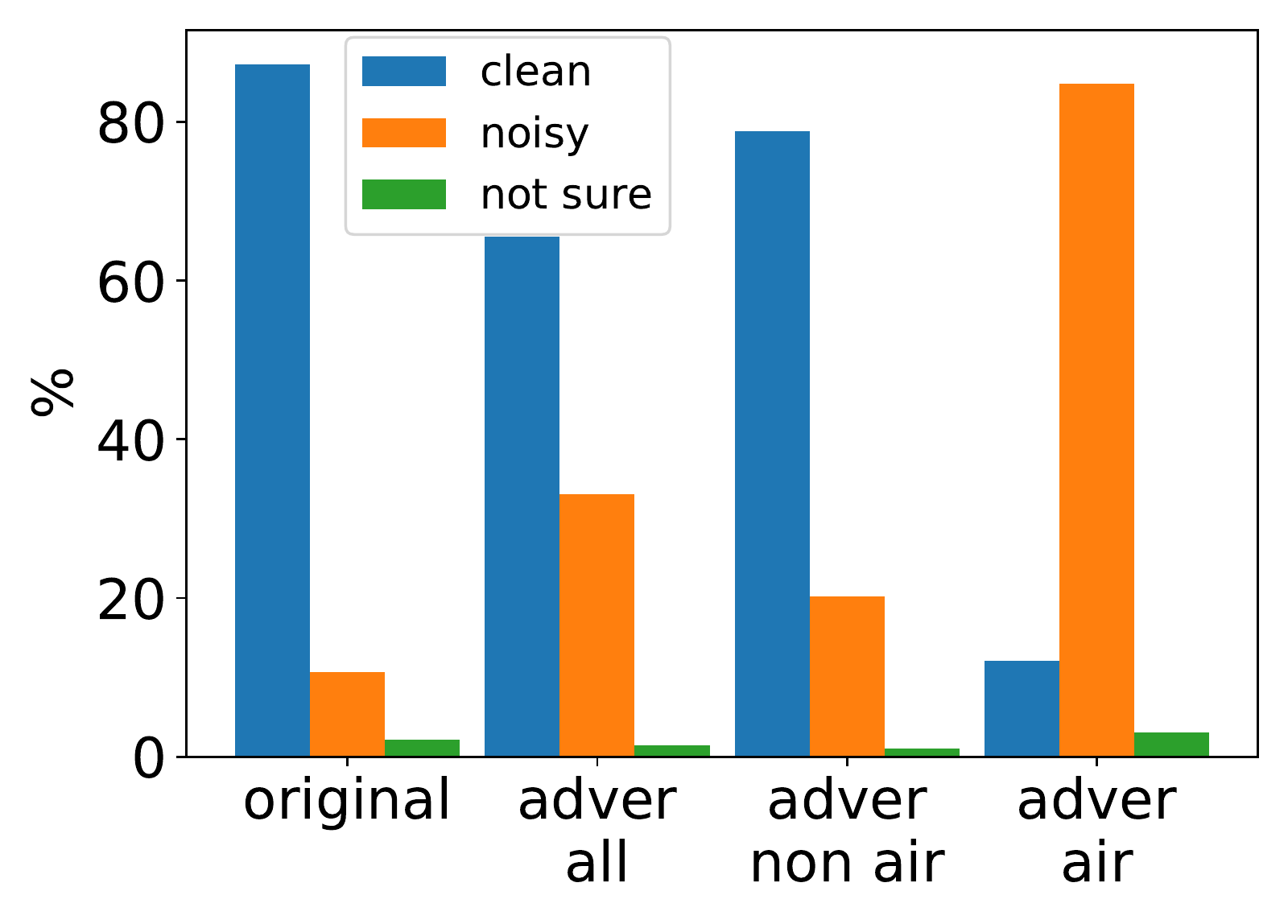}
		\caption{Task 1: clean or noisy}
	\end{subfigure}%
	\begin{subfigure}[t]{0.25\textwidth}
		\centering
		\includegraphics[width=1\textwidth]{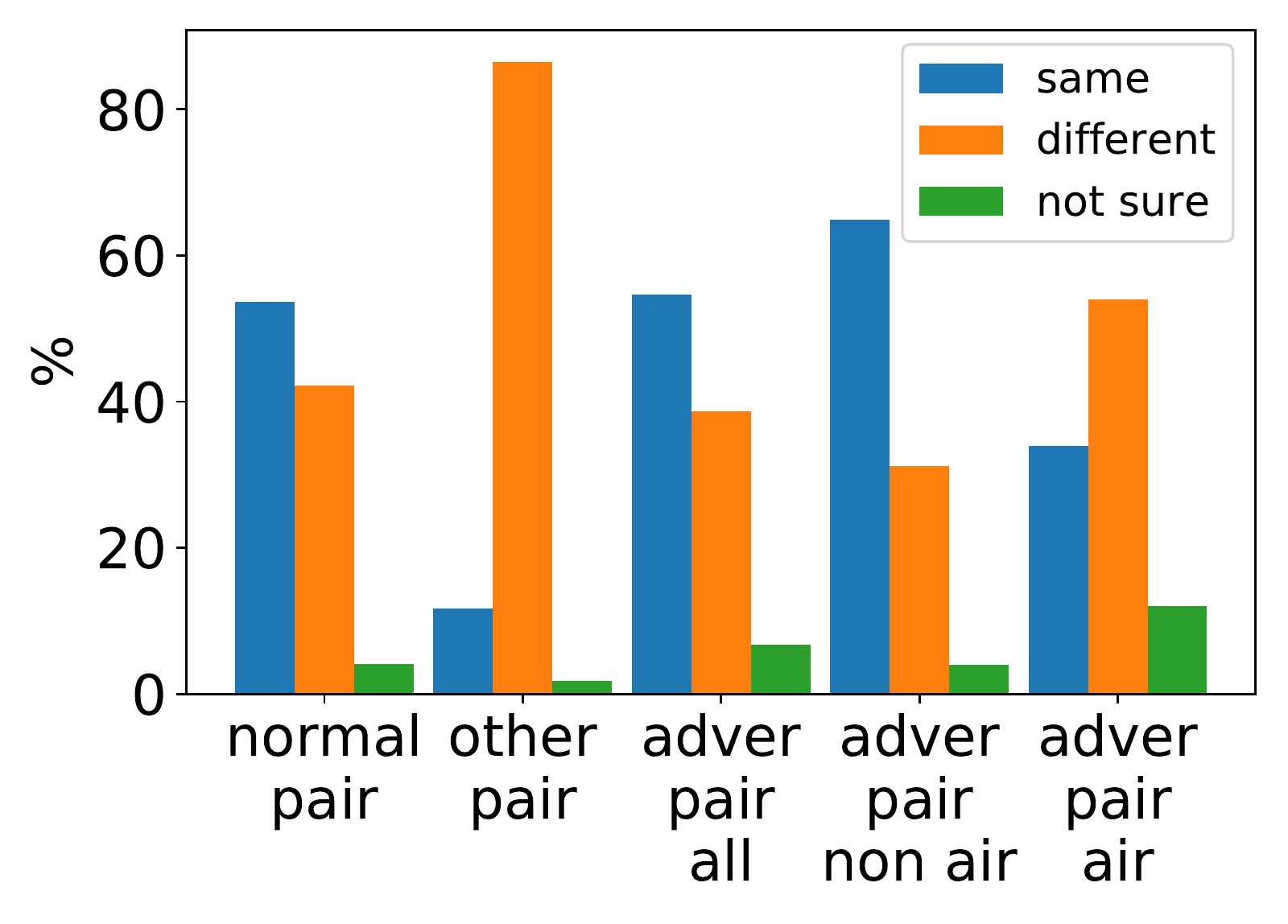}
		\caption{Task 2: identify the speaker}
	\end{subfigure}
	%\vspace{-0.5em}
	\caption{{Results of human study, where air (resp. non air) denotes voices that are effective (resp. ineffective) for over-the-air attack}}
	\label{fig:human_study_result}
	%\vspace{-2mm}
\end{figure}

\smallskip
\noindent \textbf{Results of human study.}
{We finally received 135 questionnaires for task 1 and
172 questionnaires for task 2, where 27 and 11 questionnaires are filtered out
as they failed to pass our concentration tests.
Therefore, there are 108 valid questionnaires for task 1 and
161 valid questionnaires for task 2.}
The results of the human study are shown in Fig.~\ref{fig:human_study_result}.

{For task 1, as shown in Fig.~\ref{fig:human_study_result}(a), 10.7\% of participants heard noise on normal voices, while 20.2\% and 84.8\% of participants heard noise on ineffective and effective adversarial voices (when played over-the-air) respectively.
We can see that 78.8\% of participants still believe that ineffective voices are clean.
For effective voices, we found that 84.8\% is comparable to the recent white-box adversarial attack (i.e., 83\%) that tailors to craft imperceptible voices against speech recognition systems~\cite{qin2019imperceptible}.
(We are not aware of any other adversarial attacks against SRSs that have done such human study.)}

For task 2 which is more interesting (in Fig.~\ref{fig:human_study_result}(b)),
86.5\% of participants believe that voices in each \textit{other pair} are uttered by different speakers, indicating the quality of collected questionnaires.
For the \textit{adversarial pairs},  54.6\% of participants believe that voices in each pair are uttered by the same speaker, very close
to the baseline 53.7\% of \textit{normal pairs}, indicating that humans cannot differentiate the speakers of the normal and adversarial voices.
{The prior work~\cite{kreuk2018fooling} conducted an ABX testing on adversarial samples crafted
by white-box attacks against SV systems.
The ABX test first provides to users two voices $A$ and $B$, each being either
the original (reconstructed) voice or an adversarial voice; then provides the third voice $X$ which was randomly chosen
from $\{A,B\}$; finally asks the users to decide if $X$ is $A$ or $B$.
The ABX testing of~\cite{kreuk2018fooling} shows that 54\% of participants
correctly classified the adversarial voices, which is very close to ours.}
% For ineffective voices,
% 64.9\% of participants believed that they are from the same speakers, much greater than the baseline 53.7\%, thus more imperceptible.
% {For the adversarial voices that are still effective when played over the air,
% 54.0\% of participants can definitely differentiate them, not too larger than the baseline 42.2\%.
{
For the \textit{adversarial pairs} which contain ineffective adversarial voices,
64.9\% of participants believed that the two voices are from the same speakers, much greater than the baseline 53.7\%, thus more imperceptible.
For the \textit{adversarial pairs} which contain effective adversarial voices,
54.0\% of participants can definitely differentiate the speaker, not too larger than the baseline 42.2\% of \textit{normal pairs}.
}

{The results unveil that
the adversarial voices crafted by \attackname can make systems misbehave (i.e., making a decision that the adversarial voice is uttered by the target speaker),
while most of ineffective adversarial samples are classified clean and cannot be differentiated by ordinary users,
and the results of effective ones are comparable to existing related works.
Hence, our attack is reasonably human-imperceptible.}

%%%%%%%%%%%%%%%%%%%%%%%%%%%%%%%%%%%%%%%%%%%%%%%%%%%%%%%%%%%%%%%%%

\subsection{Robustness of \attackname against Defense Methods}\label{sec:defense}
{As mentioned in Section~\ref{sec:designphi}, we study four defense methods:
local smoothing, quantization, audio squeezing and temporal dependency detection.
We evaluate on the OSI task of the GMM system unless explicitly stated
using 100 seed voices.
The FRR, FAR, ASR and UTR of the system without defense is 4.2\%, 11.2\%, 99\% and 99\%, respectively.
We consider two settings:
(S1) crafting adversarial voices on the system without defense
and attacking the system with defense,
and (S2) directly attacking the system with defense.
S1 follows from CommanderSong~\cite{yuan2018commandersong}.
An effective defense method should be able to mitigate the perturbation or detect the adversarial voices in S1.
Thus, we will use the UTR metric.
In S2, an effective defense method should increase the overhead of the attack and decrease the attack success rate,
thus we will use the ASR metric.
We set $\epsilon=0.002$, a very weak attacker capacity.
Increasing $\epsilon$ will make \attackname more powerful.}

{We found that the local smoothing can increase attack cost, but is ineffective in terms of ASR,
audio squeezing is ineffective in terms of both attack cost and ASR,
while the other two are not suitable for defending our attack.
Due to space limitation, details are given in Appendix~\ref{sec:defense}.}

%% file: discussion_future_work.tex
% !TeX root = ../main.tex

\section{Discussion of the Possible Arm Race}\label{sec:discussion_future_work}
This section discusses the  potential mitigation of our attacks and possible advanced attacks.

\smallskip
\noindent{\bf Mitigation of \attackname.}
We have demonstrated that {four defense methods have limited effects on  \attackname although some of them are reported promising in the speech recognition domain.} This reveals that more effective defense methods are needed to mitigate \attackname.
We discuss several possible defense methods as follows.

Various liveness detection methods have been proposed to detect spoofing attacks on SRSs.
Such methods detect attacks by exploiting the different physical characteristics of the voices generated by the human speech production system (i.e., lungs, vocal cords, and vocal tract) and electronic loudspeaker.
For instance, Shiota et al.~\cite{SVYOEM15} use pop noise caused by human breath,
VoiceLive~\cite{Zhang2016} leverages time-difference-of-arrival of voices to the receiver,
and VoiceGesture~\cite{zhang2017hearing} leverages the unique articulatory gesture of the user.
Adversarial voices also need to be played via loudspeakers, hence liveness detection could be possibly used to detect them.
An alternative detection method is to train
a detector using adversarial voices and normal voices.
Though promising in image recognition domain~\cite{gong2017adversarial},
it has a very high false-positive rate and does not improve the robustness when the adversary is aware of this defense~\cite{carlini2017adversarial}.
Another scheme to mitigate adversarial images is input transformation such as image bit-depth reduction and JPEG compression~\cite{guo2017countering}.
We could mitigate adversarial voices by leveraging input transformations such as bit-depth reduction and MP3 compression.
However, Athalye et al.~\cite{athalye2018obfuscated} have demonstrated that input transformation on images can be easily circumvented by strong attacks such as Backward Pass Differentiable Approximation.
We conjecture that bit-depth reduction and MP3 compression may become ineffective for high-confidence adversarial voices.

Finally, one could also improve the security
of SRSs by using a text-dependent system and requiring users to read dynamically and randomly generated sentences.
By doing so, the adversary has to attack both the speaker recognition and the
speech recognition, hence incurring attack costs.
{If the set of phrases to be uttered is relatively small, we could also attack the system
by iteratively querying the target system using the voice corresponding to the generated phrase.
While our attack will fail when the set of phrases to be uttered is very large or even infinite.
However, this also brings the challenge for the recognition system,
as the training data may not be able to cover all the possible normal phrases and voices.
%as the train data may not be able to cover all the possible normal phrases and voices.
}

In our future work, we will study the above methods~\cite{SVYOEM15, Zhang2016, zhang2017hearing,
guo2017countering,
athalye2018obfuscated, du2019deepstellar,zhang2020towards} for adversarial attacks.
%In the rest of this section, 
We next discuss possible methods on improving
adversarial attacks.

\smallskip
\noindent{\bf Possible advanced attacks.}
For a system that outputs the decision result and scores, \attackname can directly craft adversarial voices via interacting with it.
However, for a system that only outputs the decision result,
we have to attack it by leveraging transferability.
When the gap between source and target systems is larger, the transferability rate is limited.
One possible solution to improve \attackname is to leverage the boundary attack, which is proposed to attack
decision-only image recognition systems by Brendel et al.~\cite{brendel2017decision}.

{Our human study shows that our attack is reasonably human-imperceptible.
However, many of effective adversarial voices are still noisier than original voices (human study task 1),
and some of effective adversarial voices can be differentiated from
different speakers by ordinary users (human study task 2),
there still has space for improving imperceptibility in future.
One possible solution is to build a
psychoacoustic model and limit the maximal difference between the spectrum of the original and adversarial voices to the masking threshold (hearing threshold) of human perception~\cite{schnherr2018adversarial,qin2019imperceptible}.
}

% that adversarial voices generated by \attackname is imperceptible (i.e., users cannot differentiate the speakers of the original and
%adversarial voices),
%indicating that restricting the maximal value of $L_\infty$ is feasible and usually sufficient to
%craft imperceptible adversarial voices.
%However, some of the adversarial voices are still noisier than original voices (human study task 1).
%One possible solution to improve imperceptibility is to build a
%psychoacoustic model and limit the maximal difference between the spectrum of the original and adversarial voices to the %masking threshold (hearing threshold) of human perception~\cite{schnherr2018adversarial,qin2019imperceptible}.

%% file: related_work.tex
% !TeX root = ../main.tex
\vspace{1mm}
\section{Related Work}\label{sec:related_work}
%\vspace{1mm}
The security issues of intelligent voice systems have been studied in the literature.
In this section, we discuss the most related work on attacks over the intelligent voice systems, and compare them with \attackname.

\smallskip
\noindent {\bf Adversarial voice attacks.}
Gong et al.~\cite{gong2017crafting} and Kreuk et al.~\cite{kreuk2018fooling}
respectively proposed adversarial voice attacks on SRSs in the white-box setting, by leveraging the Fast Gradient Sign Method (FGSM)~\cite{goodfellow2014explaining}.
{The attack in~\cite{gong2017crafting} addresses DNN-based gender recognition, emotion recognition  and CSI systems,
while the attack in~\cite{kreuk2018fooling} addresses a DNN-based SV system.
Compared to them: (1) Our attack \attackname is black-box and more practical.
(2) \attackname addresses not only the SV and CSI, but also
the more general OSI task.
(3) We demonstrate our attack on ivector, GMM and DNN-based systems in the popular open-source platform Kaldi.
(4) \attackname is effective on the commercial systems, even when playing over the air,
which was not considered in~\cite{gong2017crafting,kreuk2018fooling}.}

In a concurrent work, Abdullah et al.~\cite{ARGBWYST19} proposed a poisoning attack on speaker and speech recognition systems,
that is demonstrated on the {OSI-Azure}.
There are three key differences:
{(1) Their attack crafts an adversarial voice from a voice uttered by an \emph{enrolled} speaker $A$
such that the adversarial voice is neither rejected nor recognized as the speaker $A$.
Thus, their attack neither can choose a specific source speaker nor a specific target speaker to be recognized by the system,
consequently, they cannot launch targeted attack or attacks against the SV task.
Whereas our attack goes beyond their attack.}
(2) They craft adversarial voice by decomposing and reconstructing an input voice,
hence, achieved a limited untargeted success rate and cannot be adapted to launch more interesting and powerful targeted attacks.
(3) We evaluate over-the-air attacks in the physical world, but they did not.

We cannot compare the performance (i.e., effectiveness and efficiency) of our attack with the three related works above~\cite{gong2017crafting,kreuk2018fooling,ARGBWYST19} because all of them are not available.
{We are the first considering the threshold $\theta$ in adversarial attack.}
Adversarial attacks on speech recognition systems also have been studied~\cite{taori2018targeted,carlini2018audio,alzantot2018did}.
Carlini et al.~\cite{carlini2018audio} attacked DeepSpeech~\cite{HCCCDEPSSCN14}
by crafting adversarial voices in the white-box setting, but failed to attack when playing over the air.
In the black-box setting, {Rohan et al.~\cite{taori2018targeted}} combined a genetic algorithm with finite difference gradient estimation to craft adversarial voices for DeepSpeech, but achieved a limited success rate with strict length restriction over the voices.
Alzantot et al.~\cite{alzantot2018did} presented the first black-box adversarial attack on a CNN-based speech command classification model by exploiting a genetic algorithm.
However, due to the difference between speaker recognition and speech recognition, these works are orthogonal to our work
and cannot be applied to ivector and GMM based SRSs.

\smallskip
\noindent {\bf Other types of voice attacks.}
Other types of voice attacks include hidden voice attack (both against speech and speaker recognition) and spoofing attack (against speaker recognition).

Hidden voice attack aims to embed some information (e.g., command) into an audio carrier (e.g., music)
such that the desired information is recognized by the target system without catching victims' attention.
Abdullah et al.~\cite{abdullah2019practical} proposed such an attack on speaker and speech recognition systems.
There are two key differences:
(1) Based on characteristics of signal processing and psychoacoustics,
their attack perturbed a sample uttered by an \emph{enrolled speaker} such that
it is still correctly classified as the \emph{enrolled speaker} by the target system but becomes incomprehensible to human listening.
While our attack perturbed a sample uttered by an \emph{arbitrary speaker}
 such that it is misclassified as a
target speaker (targeted attack) or another enrolled speaker
(untargeted attack) but the perturbation is imperceptible to human listening.
This means their attack addresses a different attack scenario compared with ours.
(2) They did not demonstrate over-the-air attack on SRSs
and their tool is not available, hence it is unclear how effective it is on SRSs.
DolphinAttack~\cite{zhang2017dolphinattack},
CommanderSong~\cite{yuan2018commandersong} and the work done by {Carlini  et al.~\cite{carlini2016hidden}}  proposed hidden voice attacks on SRSs.
{Carlini et al.} launched both black-box (i.e., inverse MFCC) and white-box (i.e., gradient decent)
attacks on GMM based speech recognition systems.
DolphinAttack exploited vulnerabilities of microphones and employed the ultrasound as the carrier of commands
to craft inaudible voices. However, it can be easily defended by filtering out the ultrasound from voices.
CommanderSong launched white-box attacks by exploiting a gradient descent method
to embed commands into music songs.

%Another attack type on SRSs is spoofing attack~\cite{wu2015spoofing}
%such as mimic~\cite{hautamaki2013vectors}, replay~\cite{wu2014study}, voice synthesis~\cite{de2012evaluation}, and voice conversion~\cite{wu2013voice} attacks.
%Spoofing attack aims at obtaining
%a voice (e.g., record-and-replay) such that
%it is correctly classified as the target speaker by the system, and also sound like the \emph{target speaker} listened by ordinary users.
%The victim will hear her/his own voice in spoofing attack,
%but hear the source speaker's voice in our attack.
%When the victim can hear the voice used for attack,
%our attack is less likely to catch the victim's attention,
%thus more practical and surreptitious.
{Another attack type on SRSs is spoofing attack~\cite{wu2015spoofing}
such as mimic~\cite{hautamaki2013vectors}, replay~\cite{wu2014study,shirvanian2019quantifying}, recorder attack~\cite{shirvanian2014wiretapping,shirvanian2019quantifying},
voice synthesis~\cite{de2012evaluation}, and voice conversion~\cite{wu2013voice,mukhopadhyay2015all,shirvanian2018short,shirvanian2019quantifying} attacks.
Different from adversarial attack~\cite{kreuk2018fooling,chen2018automated}, spoofing attack aims at obtaining a voice such that
it is correctly classified as the target speaker by the system, and also sound like the \emph{target speaker} listened by ordinary users.
When anyone familiar with the victim (including the victim) cannot hear the attack voice,
both spoofing and adversarial attacks can be launched.
However, if someone familiar with the victim (including the victim) can hear the attack voice,
he/she may detect the spoofing attack.
Whereas, adversarial attack could be launched in this setting as discussed in Section~\ref{subsec:threat}.}

%% file: conclusion.tex
% !TeX root = ../main.tex

%%%%%%%%%%%%%%%%%%%%%%%%%%%%%%%%%%%%%%%%%%%%%%%%%%%%%%%%%%%%%%%%%
\vspace{2mm}
\section{Conclusion}\label{sec:Conclusion}
\vspace{2mm}
In this paper, we conducted the first comprehensive and systematic study of adversarial attack on SRSs in a practical black-box setting, by proposing a novel practical adversarial attack \attackname.
\attackname was thoroughly evaluated in 16 attack scenarios.
{\attackname can achieve 99\% targeted attack success rate on both open-source and the commercial systems.
We also demonstrated the transferability of \attackname on Microsoft Azure.
When played over the air in the physical world, \attackname is also effective.}
Our findings reveal the security implications of \attackname for SRSs,
calling for more robust defense methods to better secure SRSs against such practical adversarial attacks.

%% file: appendices.tex
% !TeX root = ../main.tex

\appendix

\subsection{Comparison of our \attackname and PSO-based Method}\label{sec:psocomparison}
W compare our attack \attackname over a PSO-based method.
We reduce the finding of an adversarial sample as an optimization problem (cf. \S\ref{sec:problemformal}),
then solve the optimization problem via the PSO algorithm.
PSO solves the optimization problem by imitating the behaviour of a swarm of birds~\cite{eberhart1995new}. Each particle is a candidate solution, and in each iteration, the particle updates itself by the weighted linear combination of three parts, i.e.,  inertia, local best solution and global best solution.
The related weights are initial inertia factor $w_{init}$, final inertia factor $w_{end}$, acceleration constant $c_1$ and $c_2$.

\begin{table*}\centering\setlength{\tabcolsep}{4pt}\renewcommand{\arraystretch}{1.1}
	\caption{Our attack \attackname vs. the PSO-based method, where
 $[S(x_0)]_t$ denotes the initial score of input voice of the speaker $t$, and $\ast$ denotes that only one adversarial attack succeeds.}
 \vspace{-2mm}
	\label{tab:comparison_pso}
		\begin{tabular}{c|c|c|c|c|c|c|c|c|c|c|c|c}
			\hline
			\multirow{2}{*}{} & \multicolumn{2}{c|}{\textbf{$-\infty<[S(x_0)]_t<\infty$}} & \multicolumn{2}{c|}{\textbf{$[S(x_0)]_t\leq -0.5$}} & \multicolumn{2}{c|}{\textbf{$-0.5<[S(x_0)]_t\leq 0$}} & \multicolumn{2}{c|}{\textbf{$0<[S(x_0)]_t\leq 0.5$}} & \multicolumn{2}{c|}{\textbf{$0.5<[S(x_0)]_t\leq 1$}} & \multicolumn{2}{c}{\textbf{$1<[S(x_0)]_t\leq 1.5$}} \\ \cline{2-13}
			& \textbf{\attackname} & \textbf{PSO} & \textbf{\attackname} & \textbf{PSO} & \textbf{\attackname} & \textbf{PSO$^\ast$} & \textbf{\attackname} & \textbf{PSO} & \textbf{\attackname} & \textbf{PSO} & \textbf{\attackname} & \textbf{PSO} \\ \hline
			\textbf{\#Iteration} & 86 & 136 & 187 & --- & 84 & 72 & 61 & 147 & 17 & 297 & 4 & 24 \\ \hline
			\textbf{Time (s)} & 2277 & 2524 & 4409 & --- & 1947 & 1311 & 1384 & 2715 & 357 & 5517 & 77 & 449 \\ \hline
			\textbf{SNR (dB)} & 31.5 & 31.9 & 31.4 & --- & 30.5 & 22.8 & 31.5 & 31.6 & 32.4 & 32.3 & 31.8 & 32.2 \\ \hline
			\textbf{ASR (\%)} & 99.0 & 33.0 & 96.3 & 0.0 & 100.0 & 5.3 & 100.0 & 17.6 & 100.0 & 60.0 & 94.1 & 100.0 \\ \hline
		\end{tabular}
\end{table*}

\begin{table*}
	\centering
% 	\caption{{Experiment results of \attackname on xvector-PLDA system}}
\caption{{Experimental results of \attackname on xvector system}}
\vspace{-2mm}
	\label{tab:attack_result_xvector}\renewcommand{\arraystretch}{1.1}
	%\resizebox{\textwidth}{!}{%
	 \begin{tabular}{c|c|c|c|c|c|c|c|c|c|c|c|c|c|c|c|c}
		\hline
		\multirow{3}{*}{\textbf{Task}} & \multicolumn{8}{c|}{\textbf{All}} & \multicolumn{4}{c|}{\textbf{Intra-gender attack}} & \multicolumn{4}{c}{\textbf{Inter-gender attack}} \\ \cline{2-17}
		& \multicolumn{4}{c|}{\textbf{Targeted Attack}} & \multicolumn{4}{c|}{\textbf{Untargeted Attack}} & \multicolumn{4}{c|}{\textbf{Targeted Attack}} & \multicolumn{4}{c}{\textbf{Targeted Attack}} \\ \cline{2-17}
		& \textbf{\#Iter} & \textbf{\begin{tabular}[c]{@{}c@{}}Time\\ (s)\end{tabular}} & \textbf{\begin{tabular}[c]{@{}c@{}}SNR\\ (dB)\end{tabular}} & \textbf{\begin{tabular}[c]{@{}c@{}}ASR\\ (\%)\end{tabular}} & \textbf{\#Iter} & \textbf{\begin{tabular}[c]{@{}c@{}}Time\\ (s)\end{tabular}} & \textbf{\begin{tabular}[c]{@{}c@{}}SNR\\ (dB)\end{tabular}} & \textbf{\begin{tabular}[c]{@{}c@{}}ASR\\ (\%)\end{tabular}} & \textbf{\#Iter} & \textbf{\begin{tabular}[c]{@{}c@{}}Time\\ (s)\end{tabular}} & \textbf{\begin{tabular}[c]{@{}c@{}}SNR\\ (dB)\end{tabular}} & \textbf{\begin{tabular}[c]{@{}c@{}}ASR\\ (\%)\end{tabular}} & \textbf{\#Iter} & \textbf{\begin{tabular}[c]{@{}c@{}}Time\\ (s)\end{tabular}} & \textbf{\begin{tabular}[c]{@{}c@{}}SNR\\ (dB)\end{tabular}} & \textbf{\begin{tabular}[c]{@{}c@{}}ASR\\ (\%)\end{tabular}} \\ \hline
		\textbf{CSI} & 117 & 575 & 30.1 & 100.0 & 73 & 499 & 29.6 & 100.0 & 89 & 444 & 29.3 & 100 & 135 & 662 & 30.7 & 100.0 \\ \hline
		\textbf{SV} & 92 & 702 & 31.8 & 100.0 & - & - & - & - & 44 & 340 & 31.9 & 100.0 & 136 & 1035 & 31.7 & 100.0 \\ \hline
		\textbf{OSI} & 95 & 995 & 32.0 & 100.0 & 26 & 171 & 31.5 & 100.0 & 51 & 601 & 32.0 & 100.0 & 138 & 1380 & 32.0 & 100.0 \\ \hline
	\end{tabular}%
\end{table*}

We implement a PSO-based attack following the algorithm of Sharif et al.~\cite{SBBR16} which is used to fool face recognition systems.
After fine-tuning the above hyper-parameters, we conduct the experiment using the PSO-based method with 50 particles for a maximum of 35 epochs,
and we set the iteration limitation of each epoch to 30, $w_{init}$ to 0.9, $w_{end}$ to 0.1, $c_1$ to 1.4961 and $c_2$ to 1.4961.
The experiment is conducted on the ivector system for the OSI task.

The results are shown in Table~\ref{tab:comparison_pso}.
For comparison purposes, we also report the results of our attack \attackname in Table~\ref{tab:comparison_pso}.
Overall, the PSO-based method achieves 33\% targeted attack success rate (ASR), only one-third of \attackname, indicating that \attackname is much more effective than the PSO-based method.
Specifically, the PSO-based method is less effective for input voices whose initial scores are low.
\begin{itemize}
  \item When $[S(x_0)]_t \leq -0.5$, the PSO-based method fails to launch attack for all the voices.
  \item When $-0.5<[S(x_0)]_t\leq 0$ and $0<[S(x_0)]_t\leq 0.5$, the ASR is very low, i.e., 5.3\% and 17.6\%, respectively.
\end{itemize}
Whereas our attack \attackname is more effective no matter the initial scores of input voices.

In terms of efficiency,
\attackname takes less number of iterations and execution time than the PSO-based method,
except for the case $-0.5<[S(x_0)]_t\leq 0$ on which
the PSO-based method is able to launch a successful attack for one voice \emph{only}.
Specifically, the higher the initial score of the input voice is, the more efficient of our attack \attackname is compared to the PSO-based method.
For instance, when $0.5<[S(x_0)]_t\leq 1$, the number of iterations (resp. execution time) of the PSO-based method is 17 times (resp. 15 times) larger than the one of \attackname.

In summary, the experimental results demonstrate that our attack \attackname is much more effective and efficient than the PSO-based method.

\subsection{16 Attack Scenarios}\label{sec:13attacks}
All of following combinations are evaluated in this work, where D.\&S. denotes decision and scores.
{\[\footnotesize\left\{\begin{array}{c}
 \left(\begin{array}{c}
  \mbox{targeted} \\
  \mbox{untargeted}
\end{array}\right)\times \left(\begin{array}{c}
  \mbox{intra-gender} \\
  \mbox{inter-gender}
\end{array}\right)\times \mbox{API} \times
\left(\begin{array}{c}
  \mbox{OSI} \\
  \mbox{CSI} \\
  \mbox{SV}
\end{array}\right)\times  \mbox{D.\&S.} \\
+\\
 \mbox{targeted} \times\left(\begin{array}{c}
  \mbox{OSI} \\
  \mbox{CSI} \\
  \mbox{SV}
\end{array}\right) \times \mbox{API} \times   \mbox{decision-only}\\
+ \\
 \mbox{targeted} \times\left(\begin{array}{c}
  \mbox{OSI} \\
  \mbox{CSI} \\
  \mbox{SV}
\end{array}\right)\times \mbox{over-the-air} \times \mbox{D.\&S.}\\
+ \\
 \mbox{targeted} \times\mbox{OSI}\times \mbox{over-the-air}\times \mbox{decision-only}
\end{array}\right\}\]
%where D.\&S. denotes decision and scores.
}

\subsection{Results of Tuning the Parameter $\epsilon$}\label{sec:tuning}
Table~\ref{tab:tuning-epsilon} shows the results of tuning the parameter $\epsilon$  on both ivector and GMM systems for the CSI task.
To choose a suitable $\epsilon$, we need to trade off the imperceptibility and the attack cost. Smaller $\epsilon$ contributes to less perturbation (i.e, higher SNR), but also give rise to the attack cost (i.e, more iterations and execution time and lower success rate).
We found that 0.002 is a more suitable value of $\epsilon$ for two reasons: (1) compared with other $\epsilon$ values,
%0.05, 0.01, 0.005, 0.004, 0.003,
the average SNR of adversarial voices when $\epsilon=0.002$ is higher, indicating that $\epsilon=0.002$ introduces less perturbation, while the success rate of 0.002 is merely 1\% lower than that of other $\epsilon$ values. %0.05, 0.01, 0.005, 0.004 and 0.003.
(2) $\epsilon=0.001$ introduce less perturbation than $\epsilon=0.002$, but the success rate of $\epsilon=0.001$ drops to 41\% for ivector and 87\% for GMM, 58\% and 12\% lower than that of $\epsilon=0.002$. Moreover, the attack cost increases more sharply when decreasing $\epsilon$ from 0.002 to 0.001 compared with decreasing $\epsilon$ from 0.003 to 0.002. That is, the number of iterations and execution time of $\epsilon=0.002$ are 1.6 times and 1.4 times than that of $\epsilon=0.003$, while the number of iterations and execution time of $\epsilon=0.001$ are 2.2 times and 2.4 times than that of $\epsilon=0.002$.

\begin{table}[h]
\centering\setlength{\tabcolsep}{5pt}
	\caption{Results of tuning $\epsilon$ on the CSI task}
	\label{tab:tuning-epsilon}\renewcommand{\arraystretch}{1.1}
		\begin{tabular}{c|c|c|c|c|c|c|c|c}
			\hline
			\multirow{3}{*}{\textbf{$\epsilon$}} & \multicolumn{4}{c|}{\textbf{ivector}} & \multicolumn{4}{c}{\textbf{GMM}} \\ \cline{2-9}
			& \textbf{\#Iter} & \tabincell{c}{\textbf{Time}\\ (s)} & \tabincell{c}{\textbf{SNR}\\ (dB)} & \tabincell{c}{\textbf{ASR}\\ (\%)} & \textbf{\#Iter} & \tabincell{c}{\textbf{Time}\\ (s)} & \tabincell{c}{\textbf{SNR}\\ (dB)} & \tabincell{c}{\textbf{ASR}\\ (\%)}  \\ \hline
			0.05 & 18 & 422 & 12.0 & 100 & 18 & 91 & 16.7 & 100 \\ \hline
			0.01 & 23 & 549 & 16.2 & 100 & 16 & 81 & 19.1 & 100 \\ \hline
			0.005 & 44 & 1099 & 21.8 & 100 & 19 & 102 & 22.3 & 100 \\ \hline
			0.004 & 56 & 1423 & 23.8 & 100 & 21 & 104 & 24.0 & 100 \\ \hline
			0.003 & 76 & 2059 & 26.3 & 100 & 27 & 124 & 26.1 & 100 \\ \hline
			\textbf{0.002} & 124 & 2845 & 30.2 & 99 & 40 & 218 & 29.3 & 99 \\ \hline
			0.001 & 276 & 6738 & 36.4 & 41 & 106 & 551 & 35.7 & 87 \\ \hline
		\end{tabular}
\end{table}

\begin{table*}
	\centering\setlength{\tabcolsep}{3.5pt}\renewcommand{\arraystretch}{1.1}
		\caption{Details of source and target systems for transferability attacks, where DF denotes Dimension of feature, FL/FS denotes Frame length/Frame step, $\sharp$GC denotes the number of gaussian components, DV denotes Dimension of ivector (xvector), {and xvector is a DNN-based SRS from~\cite{SnyderGSMPK19}}.}
		\vspace{-2mm}
% 	\caption{Details of source and target systems for transferability attacks, where DF denotes Dimension of feature, FL/FS denotes Frame length/Frame step, $\sharp$GC denotes the number of gaussian components, DV denotes Dimension of ivector (xvector), {and xvector-PLDA is a DNN-based SRS from~\cite{SnyderGSMPK19}}.}
		\begin{tabular}{ccccccccccc}
			\hline
			\textbf{System ID}                  & \textbf{A}            & \textbf{B}                & \textbf{C}              & \textbf{D}            & \textbf{E}                    & \textbf{F}              & \textbf{G}             & \textbf{H}            & \textbf{I} & {\textbf{J}}       \\ \hline
			\textbf{Architecture}               & \textbf{GMM} & \textbf{ivector} & ivector   & ivector & ivector         & ivector   & ivector  & ivector & ivector   & {\textbf{xvector}} \\ \hline
			\textbf{Training set}                & Train-1 Set        & Train-1 Set            & \textbf{Train-2 Set} & Train-1 Set        & Train-1 Set                & Train-1 Set          & Train-1 Set         & Train-1 Set        & Train-1 Set & {Train-1 Set}                \\ \hline
			\textbf{Feature}                    & MFCC         & MFCC             & MFCC           & \textbf{PLP} & MFCC                 & MFCC           & MFCC & MFCC & \textbf{PLP} & {MFCC} \\ \hline
			\textbf{DF}             & 24$\times$3  & 24$\times$3      & 24$\times$3    & 24$\times$3  & \textbf{13$\times$3} & 24$\times$3    & 24$\times$3   & 24$\times$3  & \textbf{13$\times$3} & {\textbf{30}} \\ \hline % (no delta features)
			\textbf{FL/FS (ms)} & 25/10        & 25/10            & 25/10          & 25/10        & 25/10                & \textbf{50/10} & 25/10         & 25/10        & \textbf{50/10}      & {25/10} \\ \hline
			\textbf{$\sharp$GC} & 2048         & 2048             & 2048           & 2048         & 2048                 & 2048           & \textbf{1024} & 2048         & \textbf{1024}  & \textbf{--}    \\ \hline
			\textbf{DV}             & --      & 400            & 400          & 400          & 400                  & 400            & 400           & \textbf{600} & \textbf{600}  & {\textbf{512}}       \\ \hline
		\end{tabular}%
	%}
	\label{tab:system_info}
\end{table*}

%%%%%%%%%%%%%% re-test the performace of OSI system on different imposter set %%%%%%%%%%%%%%%%%%%%%%%%%%%%%%%%%%%%%%
\begin{table*}
	\centering
	\renewcommand{\arraystretch}{1} \setlength{\tabcolsep}{8pt}
	\caption{{The performance of the target systems C,...,J}}
	\label{tab:eight-systems-performance}
	 \begin{tabular}{c|c|c|c|c|c|c|c|c|c}
		\hline
		\multicolumn{2}{c|}{\diagbox[]{\textbf{Task}}{\textbf{System}}} & \textbf{C} & \textbf{D} & \textbf{E} & \textbf{F} & \textbf{G} & \textbf{H} & \textbf{I} & \textbf{J} \\ \hline
		\textbf{CSI} & \textbf{Accuracy} & 99.8\% & 99.4\% & 99.2\% & 99.8\% & 99.6\% & 99.8\% & 99.2\% & 99.2\% \\ \hline
		\multirow{2}{*}{\textbf{SV}} & \textbf{FAR} & 10.0\% & 9.8\% & 9.4\% & 10.0\% & 11.2\% & 9.8\% & 10.4\% & 10.2\% \\ \cline{2-10}
		& \textbf{FRR} & 1.2\% & 0.6\% & 1.6\% & 1.2\% & 0.8\% & 1.0\% & 2.2\% & 0.8\% \\ \hline
		\multirow{3}{*}{\textbf{OSI}} & \textbf{FAR} & 9.1\% & 8.8\% & 10.9\% & 9.2\% & 8.5\% & 8.1\% & 11.0\% & 7.7\% \\ \cline{2-10}
		& \textbf{FRR} & 1.4\% & 0.6\% & 1.6\% & 1.4\% & 1.2\% & 0.8\% & 2.2\% & 0.8\% \\ \cline{2-10}
		& \textbf{OSIER} & 0.0\% & 0.2\% & 0.2\% & 0.0\% & 0.2\% & 0.0\% & 0.4\% & 0.2\% \\ \hline
	\end{tabular}%
\end{table*}
%\begin{table*}
%\centering
%\renewcommand{\arraystretch}{1.2} \setlength{\tabcolsep}{8pt}
%\caption{The performance of the target systems C,...,J}
%\label{tab:eight-systems-performance}
%\begin{tabular}{c|c|c|c|c|c|c|c|c|c}
%\hline
%\multicolumn{2}{c|}{\diagbox[]{\textbf{Task}}{\textbf{System}}} & \textbf{C} & \textbf{D} & \textbf{E} & \textbf{F} & \textbf{G} & \textbf{H} & \textbf{I} & \textbf{J} \\ \hline
%\textbf{CSI} & \textbf{Accuracy} & 99.8\% & 99.4\% & 99.2\% & 99.8\% & 99.6\% & 99.8\% & 99.2\% & 99.2\% \\ \hline
%\multirow{2}{*}{\textbf{SV}} & \textbf{FAR} & 10.0\% & 9.8\% & 9.4\% & 10.0\% & 11.2\% & 9.8\% & 10.4\% & 10.2\% \\ \cline{2-10}
% & \textbf{FRR} & 1.2\% & 0.6\% & 1.6\% & 1.2\% & 0.8\% & 1.0\% & 2.2\% & 0.8\% \\ \hline
%\multirow{3}{*}{\textbf{OSI}} & \textbf{FAR} & 22.3\% & 10.9\% & 19.0\% & 14.1\% & 13.6\% & 11.4\% & 22.8\% & 14.1\% \\ \cline{2-10}
% & \textbf{FRR} & 1.4\% & 0.6\% & 1.6\% & 1.4\% & 1.2\% & 0.8\% & 2.2\% & 0.8\% \\ \cline{2-10}
% & \textbf{OSIER} & 0.0\% & 0.2\% & 0.2\% & 0.0\% & 0.2\% & 0.0\% & 0.4\% & 0.2\% \\ \hline
%\end{tabular}%
%\end{table*}

\begin{table*}\renewcommand{\arraystretch}{1.1}
\centering\setlength{\tabcolsep}{3.5pt}
\caption{{Results of transferability attack for CSI task (\%), where S denotes source system and T denotes target system.}}
\vspace{-2mm}
\label{tab:best-transferability-result-csi}
 \begin{tabular}{c|c|c|c|c|c|c|c|c|c|c|c|c|c|c|c|c|c|c|c|c}
\hline
\multirow{2}{*}{\textbf{\diagbox[]{S}{T}}} & \multicolumn{2}{c|}{\textbf{A}} & \multicolumn{2}{c|}{\textbf{B}} & \multicolumn{2}{c|}{\textbf{C}} & \multicolumn{2}{c|}{\textbf{D}} & \multicolumn{2}{c|}{\textbf{E}} & \multicolumn{2}{c|}{\textbf{F}} & \multicolumn{2}{c|}{\textbf{G}} & \multicolumn{2}{c|}{\textbf{H}} & \multicolumn{2}{c|}{\textbf{I}} & \multicolumn{2}{c}{\textbf{J}} \\ \cline{2-21}
 & \textbf{ASR} & \textbf{UTR} & \textbf{ASR} & \textbf{UTR} & \textbf{ASR} & \textbf{UTR} & \textbf{ASR} & \textbf{UTR} & \textbf{ATR} & \textbf{UTR} & \textbf{ASR} & \textbf{UTR} & \textbf{ASR} & \textbf{UTR} & \textbf{ASR} & \textbf{UTR} & \textbf{ASR} & \textbf{UTR} & \textbf{ASR} & \textbf{UTR} \\ \hline
\textbf{A} & --- & --- & 76.9 & 76.9 & {89.7} & {89.7} & 64.1 & 71.8 & 87.2 & {89.7} & 84.6 & 84.6 & 76.9 & 87.2 & 76.9 & 84.6 & 48.7 & 69.2 & 28.2 & 38.5 \\ \hline
\textbf{B} & 30.7 & 88.0 & --- & --- & 93.3 & 96.0 & \textbf{100.0} & \textbf{100.0} & \textbf{100.0} & \textbf{100.0} & \textbf{100.0} & \textbf{100.0} & 88.0 & 89.3 & \textbf{100.0} & \textbf{100.0} & 73.3 & 80.0 & 25.3 & 38.7 \\ \hline
\end{tabular}%
\end{table*}

\begin{table}[t]
	\centering
	\caption{{Results of \attackname when $\theta$ is tuned based on Equal Error Rate. The Equal Error Rate and corresponding threshold $\theta$ for ivector (resp. GMM) are 2.2\% and 1.75 (resp. 5.8\% and 0.103),
and $\epsilon=0.002$.}}
	\label{tab:attack-eer-result}
	%\resizebox{0.45\textwidth}{!}{%
	 \begin{tabular}{c|c|c|c|c|c|c|c|c}
			\hline
			\multirow{2}{*}{\textbf{Task}} & \multicolumn{4}{c|}{\textbf{ivector}} & \multicolumn{4}{c}{\textbf{GMM}} \\ \cline{2-9}
			& \textbf{\#Iter} & \textbf{\begin{tabular}[c]{@{}c@{}}Time\\ (s)\end{tabular}} & \textbf{\begin{tabular}[c]{@{}c@{}}SNR\\ (dB)\end{tabular}} & \textbf{\begin{tabular}[c]{@{}c@{}}ASR\\ (\%)\end{tabular}} & \textbf{\#Iter} & \textbf{\begin{tabular}[c]{@{}c@{}}Time\\ (s)\end{tabular}} & \textbf{\begin{tabular}[c]{@{}c@{}}SNR\\ (dB)\end{tabular}} & \textbf{\begin{tabular}[c]{@{}c@{}}ASR\\ (\%)\end{tabular}} \\ \hline
			\textbf{SV} & 120 & 2297 & 31.7 & 99.0 & 46 & 273 & 31.4 & 99.0 \\ \hline
			\textbf{OSI} & 125 & 2786 & 32.1 & 99.0 & 54 & 334 & 31.9 & 99.0 \\ \hline
		\end{tabular}%
	%}
\end{table}

\subsection{Experiment results of \attackname on xvector system}
\label{sec:attackDNN}
We demonstrate the effectiveness and efficiency of \attackname against a state-of-the-art DNN-based SRS~\cite{SnyderGSMPK19}, called xvector system, in which xvector is extracted from DNN networks.
We use the pre-trained xvector model from SITW recipe of Kaldi and
construct OSI, CSI and SV systems.
We use the same settings as in Section~\ref{sec:c-s-i-experiment}.
The baseline performance of the resulting systems is shown in Column J of Table~\ref{tab:eight-systems-performance}.
Moreover, we also conduct untargeted attacks against these systems. The results are shown in Table~\ref{tab:attack_result_xvector}. Our attack is able to achieve 100\% ASR, indicating \attackname is also effective
 and efficient against DNN-based SRSs.

\subsection{Robustness of \attackname against Defense Methods}\label{sec:defense}

\begin{figure}[h]
	\centering
	\begin{subfigure}[t]{0.25\textwidth}
		\centering
		\includegraphics[width=1\textwidth]{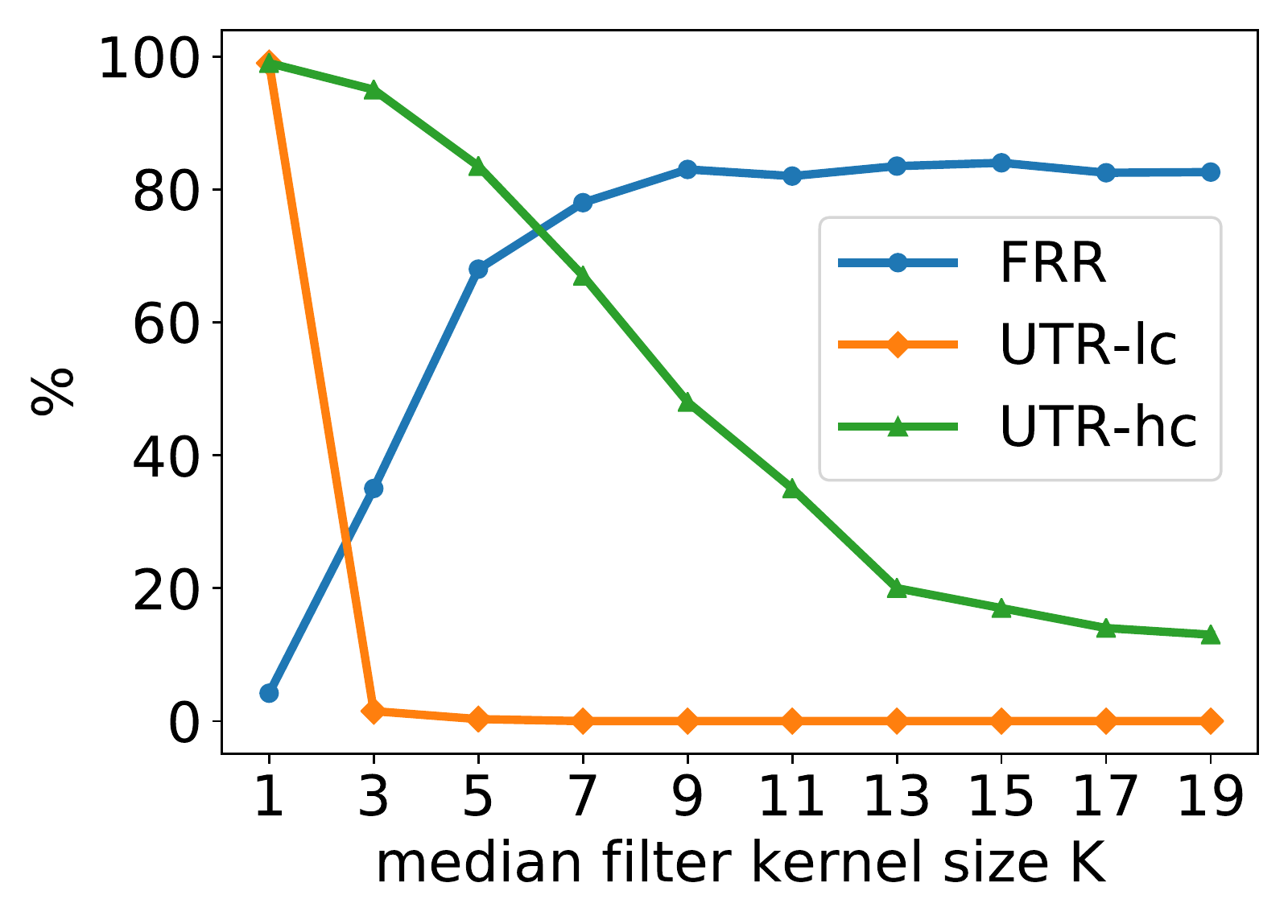}
		\caption{Median filter}
		\label{fig:median_filter}
	\end{subfigure}%
	\begin{subfigure}[t]{0.25\textwidth}
		\centering
		\includegraphics[width=1\textwidth]{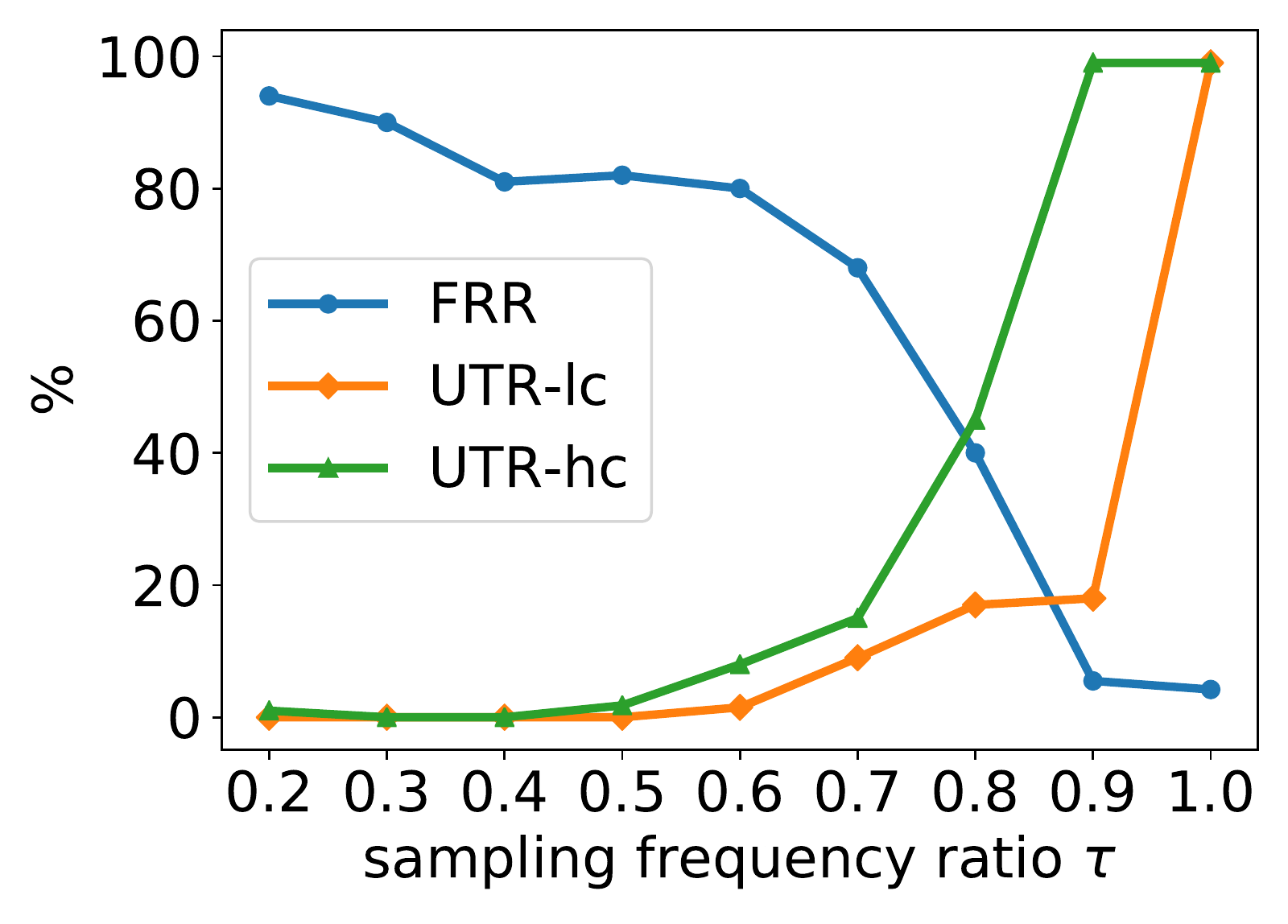}
		\caption{Audio squeezing}
		\label{fig:audio_squeezing}
	\end{subfigure}%
	\caption{{Results of median filter and audio squeezing in S1,
where UTR-lc denotes UTR of low-confidence adversarial voices ($\kappa$=0),
		and UTR-hc denotes UTR of high-confidence adversarial voices ($0<\kappa< 5$). }}
% 	\caption{{Results of median filter and audio squeezing,
% where UTR-lc denotes UTR of low-confidence adversarial voices ($\kappa$=0),
% 		and UTR-hc denotes UTR of high-confidence adversarial voices ($0<\kappa< 5$). }}
	\label{fig:defense_result}
\end{figure}

\begin{figure}[t]
	\centering
	\begin{subfigure}[t]{0.25\textwidth}
		\centering
		\includegraphics[width=1\textwidth]{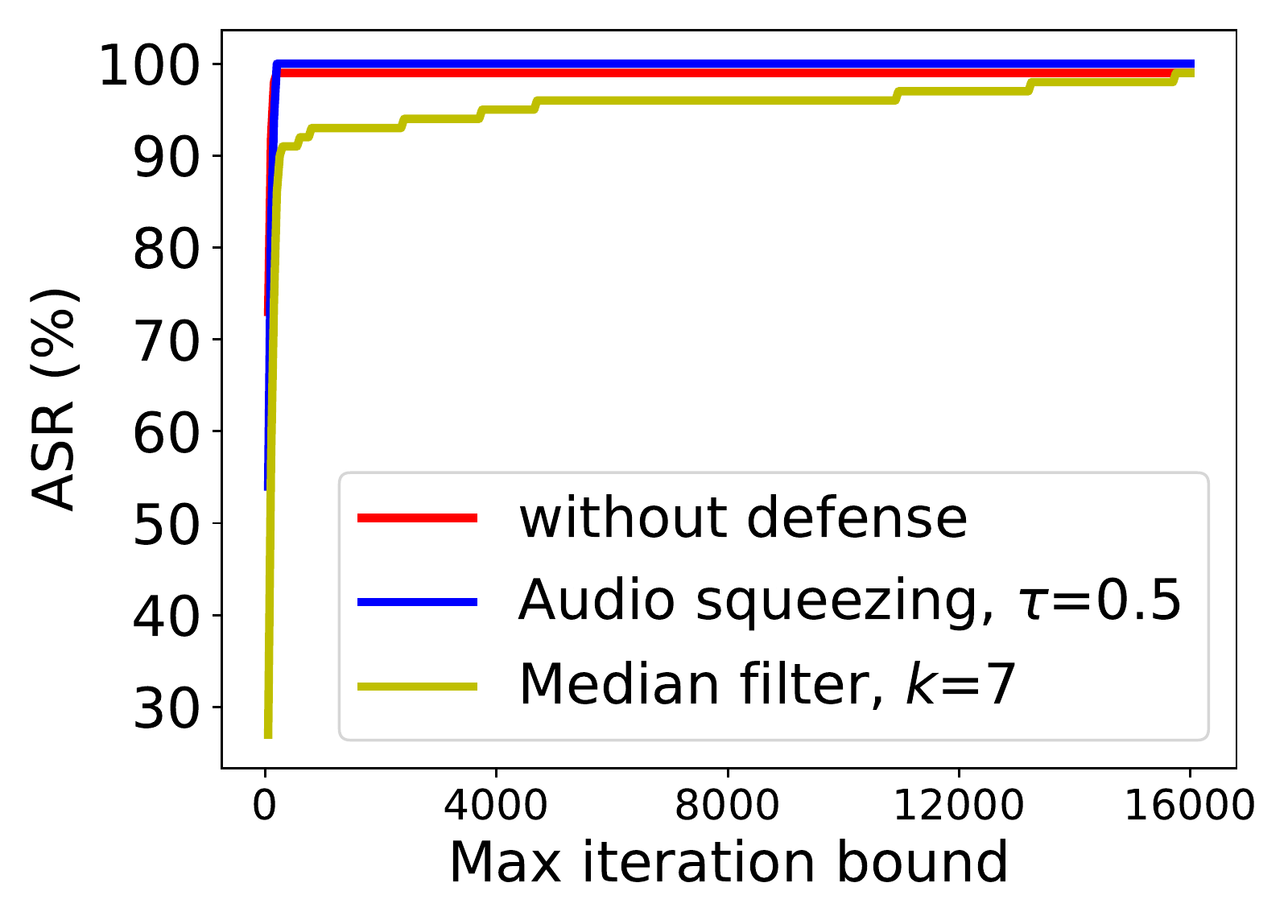}
		\caption{GMM system}
		\label{fig:gmm-defense}
	\end{subfigure}%	 	
	\begin{subfigure}[t]{0.25\textwidth}
		\centering
		\includegraphics[width=1\textwidth]{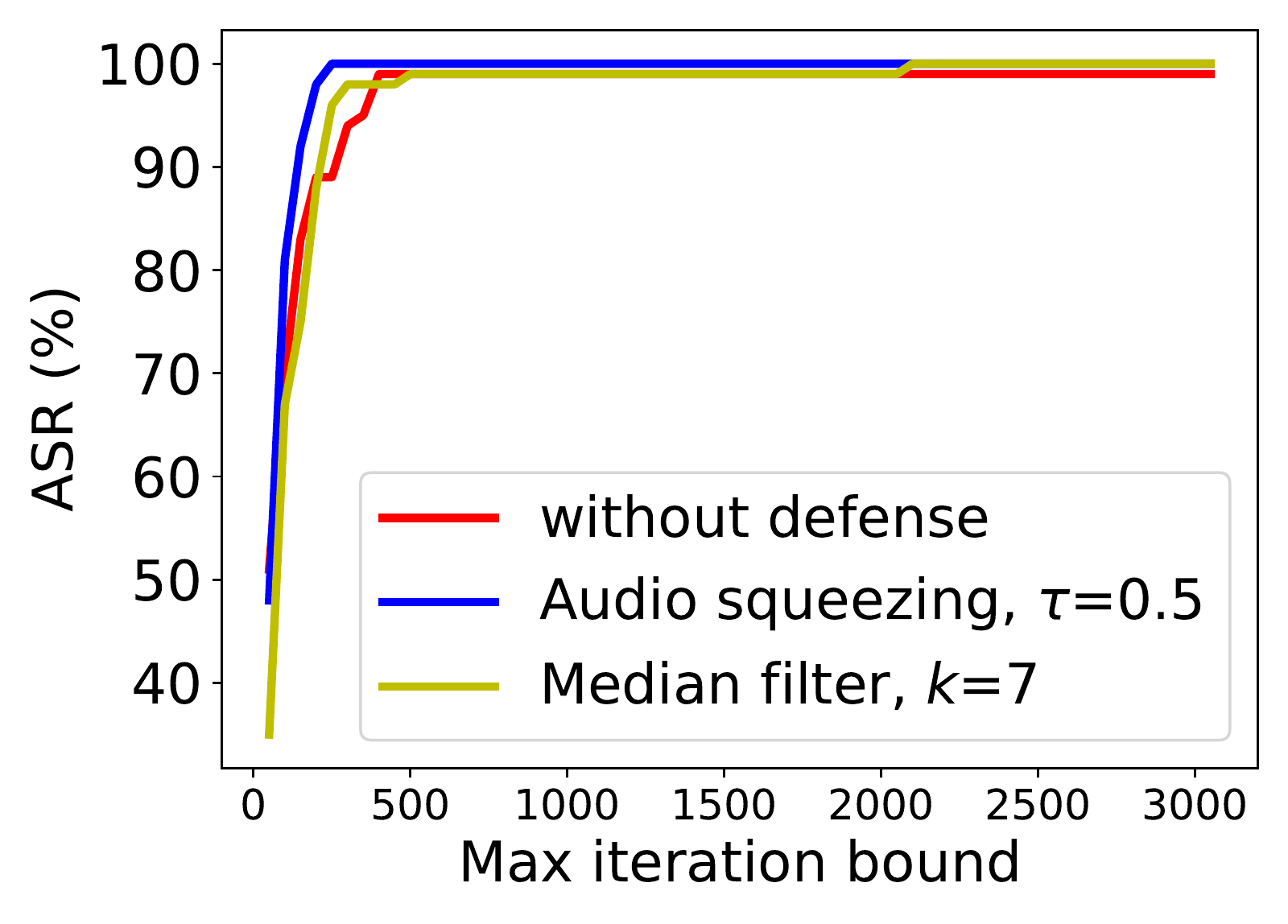}
		\caption{ivector system}
		\label{fig:iv-defense}
	\end{subfigure}%
% 	\caption{{Attack cost and ASR of median filter and audio squeezing}}
	\caption{{Attack cost of median filter and audio squzzeing}}
	\label{fig:ASR-versus-iteration}
\end{figure}

\noindent \textbf{Local smoothing.}
{It mitigates attacks by applying the mean, median or gaussian filter to the waveform of a voice.
Based on the results in~\cite{yang2018characterizing}, we use the median filter.
A median filter with kernel size $k$ (must be odd) replaces each audio element $x_k$ by the median of $k$ values [$x_{k-\frac{k-1}{2}}$, $\dots$, $x_k$, $\dots$, $x_{k+\frac{k-1}{2}}$].
In S1, we vary $k$ from 1 to 19 with step 2.
The results are shown in Fig.~\ref{fig:median_filter}.
We can see that the defense is ineffective against high-confidence (hc) adversarial voices.
For low-confidence (hc) adversarial voices,
though the UTR drops from 99\% to nearly 0\%,
the minimal FRR of normal voices increases to 35\%, significantly larger than
the baseline 4.2\%. We also tested median with $k=3$ on ivector. The FRR
of normal voices only increases by 7\%. It seems that ivector is more robust than GMM.}
{In S2, we fix $k$=7 as \cite{yang2018characterizing} did.
The results are shown in Fig.~\ref{fig:gmm-defense}.
Although the median filter increases the attack cost slightly,
\attackname can quickly achieve 90\% ASR using 250 max iteration bound, where the baseline is 90. % more than the baseline}.
To solve other few voices (9\%),  the max iteration bound should be 15,000.
Though ivector is more robust than GMM, the similar result is observed (cf. Fig.~\ref{fig:iv-defense}).}

\begin{table}[t]\renewcommand{\arraystretch}{1.2}
\centering\setlength{\tabcolsep}{3pt}
\caption{{Results of transferability attack for SV task (\%), where S: source system and T: target system.}}
\vspace{-2mm}
\label{tab:best-transferability-result-sv}
\scalebox{0.99}{ \begin{tabular}{c|c|l|c|l|c|l|c|l|c|l|c|l|c|l|c|l|c|l|c|l}
\hline
\multirow{2}{*}{\textbf{\diagbox[]{S}{T}}} & \multicolumn{2}{c|}{\textbf{A}} & \multicolumn{2}{c|}{\textbf{B}} & \multicolumn{2}{c|}{\textbf{C}} & \multicolumn{2}{c|}{\textbf{D}} & \multicolumn{2}{c|}{\textbf{E}} & \multicolumn{2}{c|}{\textbf{F}} & \multicolumn{2}{c|}{\textbf{G}} & \multicolumn{2}{c|}{\textbf{H}} & \multicolumn{2}{c|}{\textbf{I}} & \multicolumn{2}{c}{\textbf{J}} \\ \cline{2-21}
 & \multicolumn{2}{c|}{\textbf{ASR}} & \multicolumn{2}{c|}{\textbf{ASR}} & \multicolumn{2}{c|}{\textbf{ASR}} & \multicolumn{2}{c|}{\textbf{ASR}} & \multicolumn{2}{c|}{\textbf{ASR}} & \multicolumn{2}{c|}{\textbf{ASR}} & \multicolumn{2}{c|}{\textbf{ASR}} & \multicolumn{2}{c|}{\textbf{ASR}} & \multicolumn{2}{c|}{\textbf{ASR}} & \multicolumn{2}{c}{\textbf{ASR}} \\ \hline
\textbf{A} & \multicolumn{2}{c|}{---} & \multicolumn{2}{c|}{57.9} & \multicolumn{2}{c|}{49.1} & \multicolumn{2}{c|}{54.4} & \multicolumn{2}{c|}{{64.9}} & \multicolumn{2}{c|}{61.4} & \multicolumn{2}{c|}{52.6} & \multicolumn{2}{c|}{66.7} & \multicolumn{2}{c|}{36.8} & \multicolumn{2}{c}{33.3} \\ \hline
\textbf{B} & \multicolumn{2}{c|}{5.0} & \multicolumn{2}{c|}{---} & \multicolumn{2}{c|}{67.5} & \multicolumn{2}{c|}{\textbf{100.0}} & \multicolumn{2}{c|}{\textbf{100.0}} & \multicolumn{2}{c|}{\textbf{100.0}} & \multicolumn{2}{c|}{\textbf{100.0}} & \multicolumn{2}{c|}{\textbf{100.0}} & \multicolumn{2}{c|}{80.0} & \multicolumn{2}{c}{38.3} \\ \hline
\end{tabular}}%
\end{table}
\begin{table*}\renewcommand{\arraystretch}{1.1}
\centering
\caption{{Settings of the over-the-air attacks, where \emph{x meter (y dB)} means when the microphone is kept x meters away from the loudspeaker, the average volume of voices reaches y dB, and
\emph{white noise (z dB)} means the acoustic environment is degraded with a white-noise generator playing at z dB.}}
\vspace{-2mm}
\label{tab:over-the-air-four-expers}
\resizebox{1\textwidth}{!}{%
 \begin{tabular}{c|c|c|c|c|c}
\hline
  & \textbf{System} & \textbf{Loudspeaker} & \textbf{Microphone}  & \textbf{Distance} & \textbf{Acoustic Environment} \\ \hline
\textbf{\begin{tabular}[c]{@{}c@{}}Different\\ Systems \end{tabular}} & \textbf{\begin{tabular}[c]{@{}c@{}}GMM OSI/CSI/SV\\ ivector OSI/CSI/SV\\ Azure OSI\end{tabular}} & JBL clip3 portable speaker & IPhone 6 Plus (iOS) & 1 meter (65 dB) & relatively quiet \\ \hline
\textbf{\begin{tabular}[c]{@{}c@{}}Different\\ Devices \end{tabular}} & ivector OSI & \textbf{\begin{tabular}[c]{@{}c@{}}DELL laptop\\ JBL clip3 portable speaker\\ Shinco brocast equipment\end{tabular}} & \textbf{\begin{tabular}[c]{@{}c@{}}IPhone 6 Plus (iOS)\\ OPPO (Android)\end{tabular}} & 1 meter (65 dB) & relatively quiet \\ \hline
\textbf{\begin{tabular}[c]{@{}c@{}}Different\\ Distances \end{tabular}} & ivector OSI & JBL clip3 portable speaker & IPhone 6 Plus (iOS) & \textbf{\begin{tabular}[c]{@{}c@{}}0.25 meter (70 dB)\\ 0.5 meter (68 dB)\\ 1 meter (65 dB)\\ 2 meters (62 dB)\\ 4 meters (60 dB)\\ 8 meters (55 dB)\end{tabular}} & relatively quiet \\ \hline
\textbf{\begin{tabular}[c]{@{}c@{}}Different\\ Acoustic \\ Environments \end{tabular}} & ivector OSI & JBL clip3 portable speaker & IPhone 6 Plus (iOS) & 1 meter (65 dB) & \textbf{\begin{tabular}[c]{@{}c@{}}white noise (45/50/60/65/75 dB)\\ bus noise (60 dB)\\ restaurant noise (60 dB) \\  music noise (60 dB)\\ absolute music noise (60 dB)\end{tabular}} \\ \hline
\end{tabular}%
}
\end{table*}

We conclude that the local smoothing (at least median filter) can increase attack cost,
but is ineffective in terms of ASR.

\smallskip\noindent \textbf{Audio squeezing.}
{It down-samples voices and applies signal recovery to disrupt perturbations.
In S1, we vary $\tau$ (the ratio between new and original sampling frequency) from 0.1 to 1.0, the same as~\cite{yuan2018commandersong}. The results are shown in Fig.~\ref{fig:audio_squeezing}.
We can observe that when $\tau=0.9$,
(1) the FRR of normal voices is 6\%,  close to the baseline 4.2\%,
(2) the UTR of the low-confidence adversarial voices is 17\%,  smaller than
the baseline 99\%,
(3) however, the UTR of the high-confidence adversarial voices is the same as the baseline.}
{In S2, we fix $\tau$=0.5 as \cite{yang2018characterizing} did.
The results are shown in Fig.~\ref{fig:gmm-defense} and Fig.~\ref{fig:iv-defense}.
Unexpectedly, the defense decreases the overhead of attack and increases ASR.
For instance, \attackname achieves 100\% ASR using 200 max iteration bound on the system with defense,
while can only achieve 99\% ASR even
using 16,000 max iteration bound on the unsecured system.
% For instance, \attackname on the system with defense achieves 100\% ASR using 200 max iteration bound,
% while \attackname on the system without defense can only achieve 99\% ASR even
% using 16,000 max iteration bound.
It is possibly because audio squeezing ($\tau=0.5$) sacrifices the performance of SRSs.}

%100 \% using 200 max iteration bound,
%even slightly worse than the baseline (99\% ASR using 90 max iteration bound).
%It is possibly because audio squeezing ($\tau=0.5$) sacrifices the performance of SRSs.

{We conclude that the audio squeezing
is ineffective against \attackname in terms of both attack cost and ASR.}

\smallskip\noindent \textbf{Quantization.}
It rounds the amplitude of each sample point of a voice
to the nearest integer multiple of factor $q$ to mitigate the perturbation.
{In S1,} we vary $q$ from 128, 256, 512 to 1024 as~\cite{yang2018characterizing} did.
However, the system did not output any result on adversarial and normal voices.
An in-depth analysis reveals
that all the frames of voices are regarded
as unvoiced frame by the Voice Activity Detection (VAD)~\cite{sohn1999a} component.
This demonstrates that quantization is not suitable for defending against \attackname.
{Due to this, we do not consider S2.}

\smallskip\noindent {\bf Temporal dependency Detection.}
{For a given voice $v$, suppose a speech-to-text system produces text $t(v)$.
Given a parameter $0\leq k\leq 1$, let $v_k$ (resp. $t_k$) denote the
$k$ percent prefix of the voice $v$ (resp. text $t$).
The temporal dependency detection uses the distance between the texts
$t(v)_k$ and $t(v_k)$ to determine whether $v$ is an adversarial voice,
as the distance of adversarial voices is greater than that of normal voices.
We use this method to check adversarial voices crafted by \attackname using $k$=$\frac{4}{5}$ and the Character Error Rate distance metric, the best one in~\cite{yang2018characterizing}.
We do not test different values of $k$ as the result will not vary too much as mentioned in~\cite{yang2018characterizing}.
We use Baidu's DeepSpeech model as the speech-to-text system, which is implemented by Mozilla on Github~\cite{DeepSpeech-mozilla} with more than 13k stars.
}

\begin{figure}[h]
    \centering
    \includegraphics[width=0.3\textwidth]{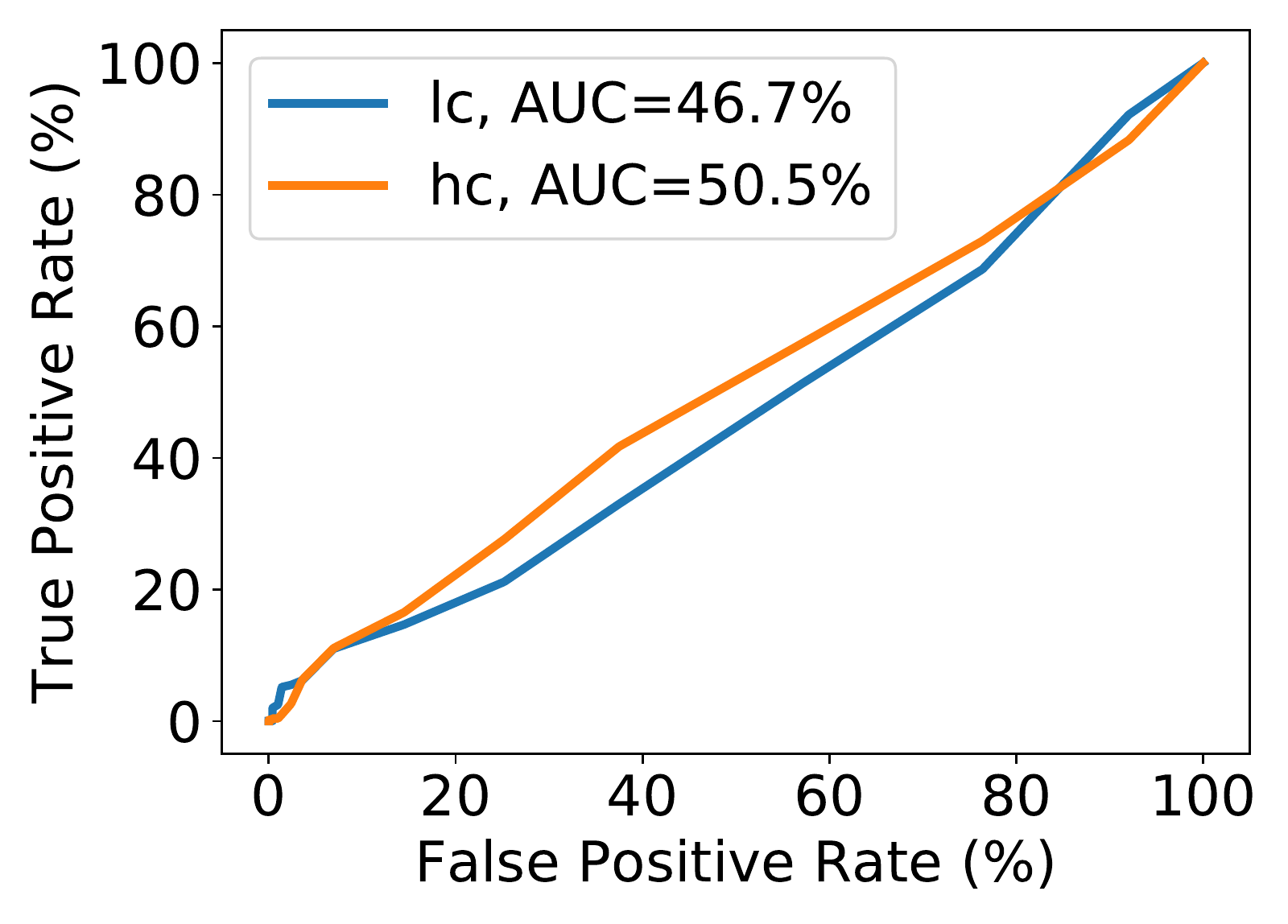}
    \caption{{ROC curves of Temporal Dependence Detection}}
    \label{fig:temporal-detection-roc}
\end{figure}

{Fig.~\ref{fig:temporal-detection-roc} shows the ROC curves of this method distinguishing low-confidence
and high-confidence adversarial samples.
It obtains 50\% true positive rate at about 50\% false positive rate.
The AUC values are 46.7\% and 50.5\%, close to random guess, indicating it fails to detect adversarial samples.
This is because \attackname does not alter the transcription of the voices, thus the temporal dependency is preserved. Due to this, we do not consider S2.}